\documentclass[reprint,superscriptaddress,amsmath,amssymb,aps,pre]{revtex4-1}

\usepackage{amsmath}
\usepackage{graphicx}
\usepackage{dcolumn}
\usepackage{bm}
\usepackage{color}
\usepackage{bbold}
\usepackage{soul}
\usepackage{subfigure}
\newcommand{\newc}{\newcommand}
\newc{\beq}{\begin{equation}}
\newc{\eeq}{\end{equation}}
\newc{\kt}{\rangle}
\newc{\br}{\langle}
\newc{\beqa}{\begin{eqnarray}}
\newc{\eeqa}{\end{eqnarray}}
\newc{\longra}{\longrightarrow}

\makeatletter
\let\Hy@backout\@gobble
\makeatother

\begin{document}

\title{Exact relations between homoclinic and periodic orbit actions in chaotic systems}

\author{Jizhou Li}
\affiliation{Department of Physics and Astronomy, Washington State University, Pullman, Washington 99164-2814, USA}
\author{Steven Tomsovic}
\affiliation{Department of Physics and Astronomy, Washington State University, Pullman, Washington 99164-2814, USA}

\date{\today}

\begin{abstract}
Homoclinic and unstable periodic orbits in chaotic systems play central roles in various semiclassical sum rules. The interferences between terms are governed by the action functions and Maslov indices. In this article, we identify geometric relations between homoclinic and unstable periodic orbits, and derive exact formulae expressing the periodic orbit classical actions in terms of corresponding homoclinic orbit actions plus certain phase space areas. The exact relations provide a basis for approximations of the periodic orbit actions as action differences between homoclinic orbits with well-estimated errors. This make possible the explicit study of relations between periodic orbits, which results in an analytic expression for the action differences between long periodic orbits and their shadowing decomposed orbits in the cycle expansion.
\end{abstract}


\maketitle

\section{Introduction}
\label{Introduction}
Phase space invariant structures, such as stable and unstable manifolds~\cite{Poincare99,Wiggins03}, play central roles in the characterization of chaotic dynamical systems. The fundamental ingredients of chaos~\cite{ChaosBook}, exponential compression and stretching, as well as the folding and mixing of phase space volumes, are all characterized by stable and unstable manifolds and the complicated patterns they form, namely homoclinic tangles~\cite{Easton86,Rom-Kedar90}. They govern various dynamical properties like phase space mixing~\cite{Ottino89,Wiggins04}, transport~\cite{Wiggins92} or escape rates~\cite{Mitchell03a,Mitchell03b,Mitchell06,Novick12a,Novick12b}. Intersections of stable and unstable manifolds give rise to homoclinic and heteroclinic orbits~\cite{Poincare99}, which have fixed past and future asymptotes.  Moreover, using regions bounded by the stable and unstable manifolds as Markov partitions~\cite{Bowen75,Gaspard98}, generic behaviors of homoclinic tangles give rise to Smale's horseshoe structures and symbolic dynamics~\cite{Smale63,Smale80}, in which the motions of points from non-wandering sets~\cite{Wiggins03,ChaosBook} under successive mappings are topologically conjugate to a Bernoulli shift on their symbolic strings~\cite{Hadamard1898,Birkhoff27a,Birkhoff35,Morse38}. 

Of particular interest to both classical and quantum chaos theory are the unstable periodic orbits and homoclinic orbits from a non-wondering set, which can be uniquely identified from the symbolic strings of the horseshoe.  For example, classical sum rules over unstable periodic orbits describe various entropies, Lyapunov exponents, escape rates, and the uniformity principle \cite{So07}. In the semiclassical regime, properties of such classical orbits are also extremely important. A few cases are given by periodic~\cite{Gutzwiller71,Balian71,Berry76} and closed orbit sum rules~\cite{Du88a,Du88b,Friedrich89} that determine quantal spectral properties, and homoclinic (heteroclinic) orbit summations~\cite{Tomsovic91b,Tomsovic93} generating wave packet propagation approximations. The interferences in such semiclassical sum rules are almost exclusively governed by the orbits' classical action functions and Maslov indices~\cite{Creagh90,Mao92,Esterlis14}, and thus this information takes on greater importance in the context of the asymptotic properties of quantum mechanics.  Various resummation techniques have been given to work with series which are often divergent in nature~\cite{Tanner91,Cvitanovic89,Berry90}. Other studies exploring a fuller understanding of the interferences have also been carried out~\cite{Argaman93,Ozorio89,Bogomolny92,Sieber01,Muller04,Turek05,Muller05}. In semiclassical orbit summations, the classical action as a phase factor, is scaled by $\hbar$. Therefore small errors in the orbit actions will be magnified and significantly compromise the degree of accuracy of the spectral quantities. Due to sensitive dependence on initial conditions, numerical methods of calculating long orbits and their classical actions become exponentially demanding with increasing periods, hindering the calculations of systems' spectra on fine-resolution scales.  

There exist rather intimate relations between orbits in chaotic systems, e.g.~similarity of two symbolic strings implies shadowing between the actual points in phase space; see, for example~\cite{Gutkin13}.  Thus, unstable periodic orbits are always shadowed by homoclinic orbit segments possessing the same symbolic codes. It turns out that periodic orbits can be built up from different homoclinic orbit segments up to any desired accuracy.  Furthermore, using the MacKay-Meiss-Percival action principle~\cite{MacKay84a,Meiss92},  exact and considerably simpler approximate formulae can be derived expressing the periodic orbit actions in terms of the homoclinic actions, which can be calculated in fast and stable ways~\cite{Li17}.  The exponential divergence problem in the numerical computations of long periodic orbits can be avoided. The same formulae also enable explicit studies of action relations between different periodic orbits, which eventually lead to analytic expressions of action differences between long periodic orbits and their decomposed pseudo-orbits in the theory of cycle expansion~\cite{Cvitanovic88,Cvitanovic89}.  

This paper is organized as follows. Sec.~\ref{Basic Concepts and Definitions} reviews the concepts related to the horseshoe map symbolic dynamics, and definitions of various kinds of generating functions in Hamiltonian systems. Sec.~\ref{Action formulae} develops the theory to express the classical actions of unstable periodic orbits as differences between selected homoclinic orbits. Sec.~\ref{Cycle expansion} demonstrates an immediate application to determine the small action differences between periodic orbits and their decomposed pseudo-orbits in the cycle expansion. Sec.~\ref{Conclusions} summarizes the work and discusses possible future research.  
  
\section{Basis Concepts and Definitions}
\label{Basic Concepts and Definitions}

\subsection{Symbolic dynamics and horseshoes}
\label{Symbolic dynamics and horseshoe}

\begin{figure}
 \subfigure{
   \label{fig:horseshoe_1}
   \includegraphics[width=5.5cm]{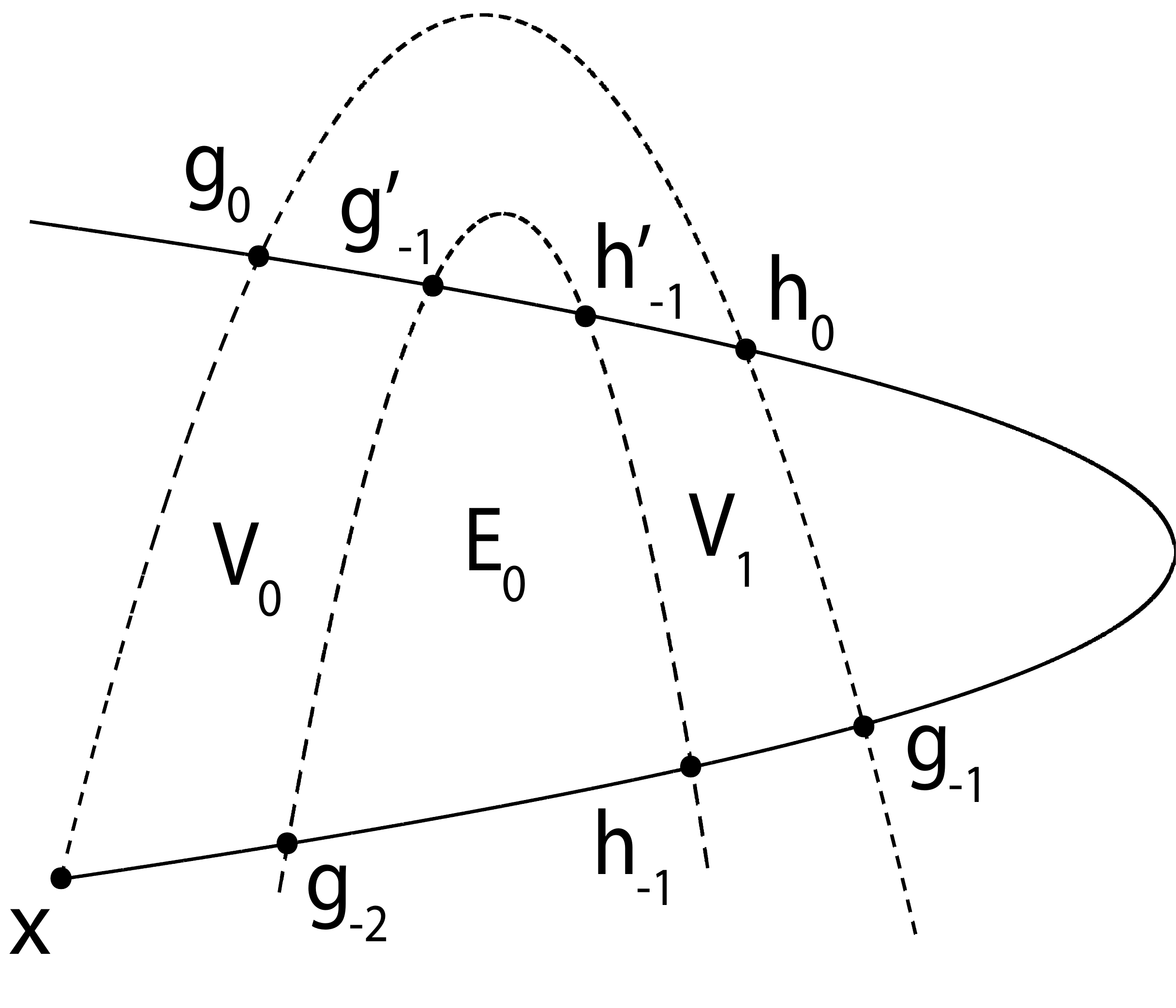}}
 \subfigure{
   \label{fig:horseshoe_2}
   \includegraphics[width=5.5cm]{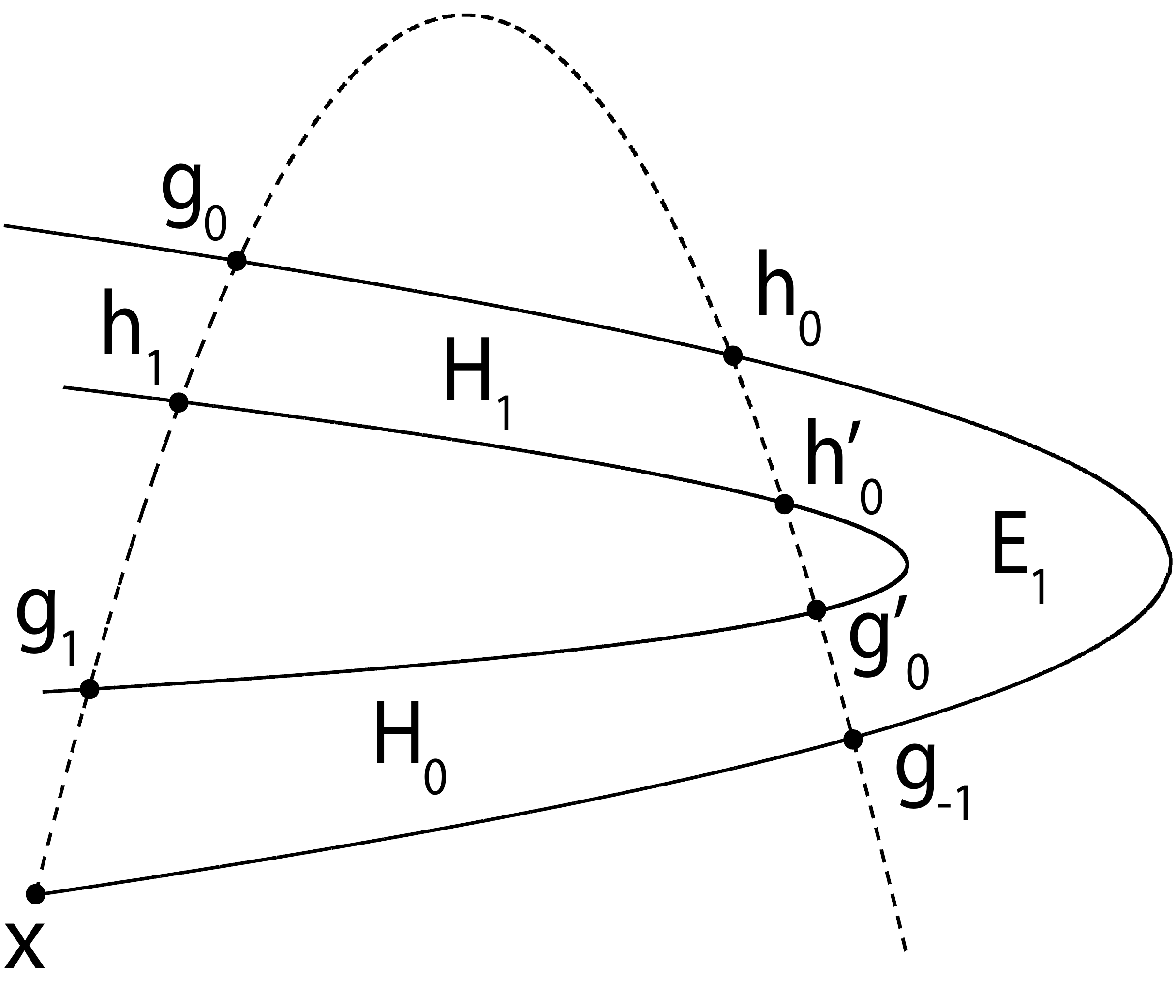}}
\caption{(Schematic) Example partial homoclinic tangle from the H\'{e}non map, which forms a complete horseshoe structure. The unstable (stable) manifold of $x$ is the solid (dashed) curve. There are two primary homoclinic orbits $\lbrace h_0\rbrace$ and $\lbrace g_0\rbrace$. $\cal{R}$ is the closed region bounded by loop $\mathcal{L}_{USUS[x,g_{-1},h_0,g_0]}$. Under forward iteration, the vertical strips $V_0$ and $V_1$ (including the boundaries) from the upper panel are mapped into the horizontal strips in the lower panel. At the same time, points in region $E_0$ are mapped outside $\cal{R}$ into region $E_1$, never to return and escape to infinity. There is a Cantor set of points in $V_0$ and $V_1$ that remain inside $\cal{R}$ for all iterations, which is the non-wandering set $\Omega$ defined in Eq.~\eqref{eq:Nonwandering set}. The phase space itineraries of points in $\Omega$ in terms of $V_0$ and $V_1$ give rise to symbolic dynamics, as described by Eq.~\eqref{eq:symbolic code}.} 
\label{fig:horseshoe}
\end{figure}

Let $M$ be an analytic and area-preserving map on the two-dimensional phase space $(q,p)$, and $x=(q,p)$ be a hyperbolic fixed point under $M$. Denote the unstable and stable manifolds of $x$ by $U(x)$ and $S(x)$, respectively. Typically, $U(x)$ and $S(x)$ intersect infinitely many times and form a complicated pattern called a homoclinic tangle~\cite{Poincare99,Easton86,Rom-Kedar90}.   The notation $U[x_1,x_2]$ is introduced to denote the finite segment of $U(x)$ extending from $x_1$ to $x_2$, both of which are points on $U(x)$, and similarly for $S(x)$. It is well-known that Markov generating partitions to the phase space \cite{Bowen75,Gaspard98} can be constructed that use segments on $U(x)$ and $S(x)$ as boundaries, which are used to assign symbolic dynamics \cite{Hadamard1898,Birkhoff27a,Birkhoff35,Morse38} as phase-space itineraries of trajectories under the mapping.  Assume the system is highly chaotic and the homoclinic tangle forms a complete horseshoe, part of which is shown in Fig.~\ref{fig:horseshoe}, as this is generic to a significant class of         
dynamical systems. In such scenarios, the Markov partition is a simple set of two regions $[V_0,V_1]$, as shown in Fig.~\ref{fig:horseshoe}. Each phase-space point $z_0$ that never escapes to infinity can be put into an one-to-one correspondence with a bi-infinite symbolic string
\begin{equation}\label{eq:symbolic code}
z_0 \Rightarrow \cdots s_{-2}s_{-1}.s_{0}s_{1}s_{2}\cdots
\end{equation}
where each digit $s_n$ in the symbol denotes the region that $M^{n}(z_0)$ lies in: $M^{n}(z_0) = z_n \in V_{s_{n}}$, $s_n \in \lbrace 0,1\rbrace$. In that sense, the symbolic code gives an ``itinerary" of $z_0$ under successive forward and backward iterations, in terms of the regions $V_0$ and $V_1$ in which each iteration lies.  Throughout this paper we use the area-preserving H\'{e}non map (Eq.~\eqref{eq:Henon map}) with parameter $a=10$ for illustration. This parameter is well beyond the first tangency, thus giving rise to a complete horseshoe-shaped homoclinic tangle with highly chaotic dynamics. It serves as a simple paradigm since the symbolic dynamics permits all possible combinations of binary codes, no ``pruning'' is needed. However, the results derived ahead mostly carry over into more complicated systems possessing incomplete horseshoes~\cite{Cvitanovic88a,Cvitanovic91}, or systems with more than binary symbolic codes, though more work is needed to address such systems. Refer to Appendix.~\ref{Markov partition} for more details on the construction of the Markov generating partition and symbolic dynamics. 

The intersections between $S(x)$ and $U(x)$ give rise to homoclinic orbits, which are asymptotic to $x$ under both $M^{\pm\infty}$. From the infinite families of homoclinic orbits, two special ones $\lbrace h_0\rbrace$ and $\lbrace g_0\rbrace$ can be identified as primary homoclinic orbits, in the sense that they have the simplest phase space excursions (the set $\lbrace h_0\rbrace$ includes the point $h_0$ and all its iterations forward and backward in time). The segments $S[x,h_0]$ and $U[x,h_0]$ intersect only at $h_0$, the same is true for all its orbit points $h_i$; this holds for $\lbrace g_0\rbrace$ as well. There are only two primary orbits for the horseshoe, but possibly more for systems with more complicated homoclinic tangles.   

The orbit of $z_0$, denoted by $\lbrace z_0 \rbrace$, is the infinite collection of all $M^{n}(z_0)$: 
\begin{equation}\label{eq:Orbit}
\begin{split}
&\lbrace z_0 \rbrace=\\
&\lbrace M^{-\infty}(z_0),\cdots,M^{-1}(z_0),z_0,M(z_0),\cdots,M^{\infty}(z_0)\rbrace \\
&=\lbrace z_{-\infty},\cdots,z_{-1},z_0,z_1,\cdots,z_{\infty} \rbrace  \nonumber
\end{split}
\end{equation}
where $z_n = M^n(z_0)$ for all $n$.  Points along the same orbit have the same symbolic strings but shifting decimal points. Therefore, an orbit can be represented by the symbolic string without the decimal point.  

Under the symbolic dynamics, a period-$T$ point $y_0$, where $M^T(y_0)=y_0$, can always be associated with a symbolic string with infinite repetitions of a substring with length $T$: 
\begin{equation}\label{eq:Periodic point}
 y_0  \Rightarrow \cdots s_0 s_1 \cdots s_{T-1} . s_0 s_1 \cdots s_{T-1} \cdots =\overline{\gamma}.\overline{\gamma}
\end{equation}
where $\gamma =s_0 \cdots s_{T-1}$ is the finite substring and $\overline{\gamma}.\overline{\gamma}$ denotes its infinite repetition (on both sides of the decimal point). Notice that the cyclic permutations of $s_0 \cdots s_{T-1}$ can be associated with the successive mappings of $y_0$, generating a one-to-one mapping to the set of points on the orbit. Since an orbit can be represented by any point on it, the position of the decimal point does not matter, therefore we denote the periodic orbit $\lbrace y_0 \rbrace$ as
\begin{equation}\label{eq:Periodic orbit}
\lbrace  y_0 \rbrace \Rightarrow \overline{\gamma} 
\end{equation}  
with the decimal point removed.  Similarly, the finite length-$T$ orbit segment $[y_0,y_1,\cdots,y_{T-1}]$, which composes one full period,  is denoted
\begin{equation}\label{eq:Periodic orbit segment}
[y_0,y_1,\cdots,y_{T-1}] \Rightarrow \gamma 
\end{equation}  
with the overhead bar removed, as compared to Eq.~\eqref{eq:Periodic orbit}.  Any cyclic permutation of $\gamma$ refers to the same periodic orbit.

A primitive periodic orbit is a periodic orbit that cannot be written into repetitions of any shorter periodic orbits. Correspondingly, its symbolic string cannot be written into repetitions of any of its shorter substrings. For example, $\overline{110110}$ is not a primitive periodic orbit, since it is just the twice mapped primitive periodic orbit $\overline{110}$\ . 
 
The hyperbolic fixed point has the simplest symbolic code $x \Rightarrow \overline{0}.\overline{0}$\ , and its orbit $\lbrace x \rbrace \Rightarrow \overline{0}$ correspondingly. A homoclinic point $h_0$ of $x$ has symbolic code of the form~\cite{Hagiwara04}: 
\begin{equation}\label{eq:Homoclinic point}
h_0 \Rightarrow \overline{0} 1 s_{-m}\cdots s_{-1}.s_0 s_1 \cdots s_n 1 \overline{0} 
\end{equation}
where the $\overline{0}$ on both ends means the orbit approaches the fixed point (therefore stays in $V_0$) under both $M^{\pm \infty}$. Similar to the periodic orbit case, the homoclinic orbit can be represented as
\begin{equation}\label{eq:Homoclinic orbit}
\lbrace h_0 \rbrace \Rightarrow  \overline{0} 1 s_{-m}\cdots s_{-1}s_0 s_1 \cdots s_n 1 \overline{0}
\end{equation}
with the decimal point removed, as compared to Eq.~\eqref{eq:Homoclinic point}. 

Of particular interest, the primary homoclinic points $g_0$ and $h_0$ in Fig.~\ref{fig:horseshoe} have symbolic codes $\overline{0}1.\overline{0}$ and $\overline{0}1.1\overline{0}$, respectively. Their forward iterations $g_1$ and $h_1$ have codes $\overline{0}10.\overline{0}$ and $\overline{0}11.\overline{0}$, respectively, which corresponds to a shift of the decimal points for one step towards to right side. The orbits $\lbrace g_0\rbrace$ and $\lbrace h_0\rbrace$ are represented by:
\begin{equation}\label{eq:Primary homoclinic orbits symbolic}
\begin{split}
\lbrace g_0 \rbrace &\Rightarrow  \overline{0}1\overline{0}  \\
\lbrace h_0 \rbrace &\Rightarrow  \overline{0}11\overline{0} \ 
\end{split}
\end{equation} 
note that they have the simplest possible codes among all homoclinic orbits. For example, the non-primary orbits $\lbrace g^{\prime}_0\rbrace$ and $\lbrace h^{\prime}\rbrace$ from Fig.~\ref{fig:horseshoe} have codes $\overline{0}101\overline{0}$ and $ \overline{0}111\overline{0}$, respectively.

\subsection{Generating function and classical action}
\label{Generating function and classical action}

For any phase space point $z_n=(q_n,p_n)$ and its image $M(z_n)=z_{n+1}=(q_{n+1},p_{n+1})$, the mapping $M$ can be viewed as a canonical transformation that maps $z_n$ to $z_{n+1}$ while preserving the symplectic area, therefore a $\mathit{generating}$ ($\mathit{action}$) $\mathit{function}$ $F(q_{n},q_{n+1})$ can be associated with this process such that~\cite{MacKay84a,Meiss92}:
\begin{equation}\label{eq:Definition generating function}
\begin{split}
&p_{n}=-\partial F/\partial q_{n}\\ 
&p_{n+1}=\partial F/\partial q_{n+1}.
\end{split}
\end{equation}
Despite the fact that $F$ is a function of $q_n$ and $q_{n+1}$, it is convenient to denote it as $F(z_n,z_{n+1})$. This should cause no confusion as long as it is kept in mind that it is the $q$ variables of $z_n$ and $z_{n+1}$ that go into the expression of $F$.  The compound mapping $M^{k}$, which maps $z_n$ to $z_{n+k}$, then has the generating function:
\begin{equation}\label{eq:Definition generating function compound map}
F(z_n,z_{n+k}) \equiv \sum_{i=n}^{n+k-1}F(z_{i},z_{i+1})
\end{equation}  
which, strictly speaking, is a function of $q_n$ and $q_{n+k}$. 

For periodic orbits $\overline{\gamma}$ with primitive period $T$, the primitive period $\mathit{classical}$ $\mathit{action}$ ${\cal F}_{\gamma}$ of the orbit is:
\begin{equation}
\label{eq:Definition generating function primitive periodic orbits}
{\cal F}_{\gamma}  \equiv \sum_{i=0}^{T-1}F(y_{i},y_{i+1})\ .
\end{equation}  
${\cal F}_{\gamma}$ is just the generating function that maps a point along the orbit for one primitive period.  For the special case of the fixed point $x$, Eq.~\eqref{eq:Definition generating function primitive periodic orbits} reduces to:
\begin{equation}\label{eq:Definition generating function fixed points}
{\cal F}_{0}  = F(x,x) 
\end{equation} 
where $F(x,x)$ is the generating function that maps $x$ into itself in one iteration.

For non-periodic orbits $\lbrace h_0 \rbrace$, the classical action is the sum of the generating functions over infinite successive mappings:
\begin{equation}
\label{eq:full orbit action in general}
{\cal F}_{\lbrace h_0 \rbrace} \equiv \lim_{N \to \infty} \sum_{i=-N}^{N-1} F(h_{i},h_{i+1})=\lim_{N \to \infty} F(h_{-N},h_{N})
\end{equation}
and is divergent in general. However, the MacKay-Meiss-Percival action principle~\cite{MacKay84a,Meiss92} can be applied to obtain well defined action differences for particular pairs of orbits.  An important and simple case is the $\mathit{relative}$ $\mathit{action}$ $\Delta {\cal F}_{\lbrace h_0 \rbrace  \lbrace x \rbrace}$ between a fixed point $x$ and its homoclinic orbit $\lbrace h_{0} \rbrace$, where $h_{\pm \infty}\to x$:
\begin{eqnarray}
\label{eq:relative action homoclinic}
\Delta {\cal F}_{\lbrace h_0 \rbrace  \lbrace x \rbrace} &\equiv & \lim_{N \to \infty} \sum_{i=-N}^{N-1}\left[F(h_i,h_{i+1})-F(x,x)\right] \nonumber \\
&=& \int\limits_{U[x,h_{0}]}p\mathrm{d}q+\int\limits_{S[h_{0},x]}p\mathrm{d}q = \oint_{US[x,h_{0}]} p\mathrm{d}q \nonumber \\
&=& {\cal A}^\circ_{US[x,h_{0}]}
\end{eqnarray}
where $U[x,h_{0}]$ is the segment of the unstable manifold from $x$ to $h_{0}$, and $S[h_{0},x]$ the segment of the stable manifold from $h_0$ to $x$.  The $\circ$ superscript on the last line indicates that the area is interior to a path that forms a closed loop, and the subscript indicates the path: $US[x,h_{0}]=U[x,h_{0}]+S[h_{0},x]$.  As usual, clockwise enclosure of an area is positive, counterclockwise negative. $\Delta {\cal F}_{\lbrace h_0 \rbrace \lbrace x \rbrace} $ gives the action difference between the homoclinic orbit segment $[ h_{-N},\cdots,h_{N} ]$ and the length-$(2N+1)$ fixed point orbit segment $[ x, \cdots, x ]$ in the limit $N \to \infty$. In later sections, upon specifying the symbolic code of the homoclinic orbit $\lbrace h_0 \rbrace \Rightarrow \overline{0} \gamma \overline{0}$, we also denote $\Delta {\cal F}_{\lbrace h_0 \rbrace  \lbrace x \rbrace}$ alternatively as
\begin{equation}\label{eq:relative action homoclinic symbolic notation}
\Delta {\cal F}_{\lbrace h_0 \rbrace  \lbrace x \rbrace} = \Delta {\cal F}_{\overline{0} \gamma \overline{0},  \overline{0}}
\end{equation}
by replacing the orbits in the subscript with their symbolic codes. 

Likewise, a second important case is for the relative action between a pair of homoclinic orbits $\lbrace h^{\prime}_0\rbrace \Rightarrow \overline{0} \gamma^{\prime} \overline{0} $ and $\lbrace h_0\rbrace \Rightarrow \overline{0} \gamma \overline{0}$, which results in 
\begin{eqnarray}
\label{eq:homoclinic action difference}
\Delta{\cal F}_{{\lbrace h^{\prime}_0\rbrace}{\lbrace h_0\rbrace}} &\equiv & \lim_{N \to \infty} \sum_{i=-N}^{N-1}\left[ F(h^{\prime}_{i},h^{\prime}_{i+1}) - F(h_{i},h_{i+1})\right] \nonumber \\
& = & \lim_{N \to \infty} \left[ F(h^{\prime}_{-N},h^{\prime}_{N}) - F(h_{-N},h_{N}) \right] \nonumber \\
& = & \int\limits_{U[h_{0},h^\prime_{0}]}p\mathrm{d}q+\int\limits_{S[h^\prime_{0},h_{0}]}p\mathrm{d}q =  {\cal A}^\circ_{US[h_0,h^\prime_{0}]} \nonumber \\
& = & \Delta{\cal F}_{ \overline{0} \gamma^{\prime} \overline{0},  \overline{0} \gamma \overline{0} } 
\end{eqnarray}
where $U[h_{0},h^\prime_{0}]$ is the segment of the unstable manifold from $h_{0}$ to $h^\prime_{0}$, and $S[h^\prime_{0},h_{0}]$ the segment of the stable manifold from $h^\prime_{0}$ to $h_{0}$.  Due to the fact that the endpoints approach $x$ forward and backward in time, one can also write
\begin{eqnarray}
\label{eq:homoclinic action difference2}
\Delta{\cal F}_{{\lbrace h^{\prime}_0\rbrace}{\lbrace h_0\rbrace}} & = & \lim_{N \to \infty} \left[ F(h^{\prime}_{-(N+n)},h^{\prime}_{N+m}) - F(h_{-N},h_{N}) \right] \nonumber \\
& & - (n+m) {\cal F}_0   \ , 
\end{eqnarray}
which is useful ahead.

\section{Action formulae}
\label{Action formulae}

In highly chaotic systems, the computation of long orbits is always a daunting task due to exponential divergence on initial error. On the contrary, invariant structures such as the stable and unstable manifolds, along with homoclinic orbits, can be calculated in rather stable ways with high precisions \cite{Li17}. It is thus desirable to extract information about periodic orbits (which are unstable to calculate) from the knowledge of homoclinic orbits (which are stable to calculate). In a previous work \cite{Li17a}, the geometric relations between the homoclinic orbits and periodic orbits were given using Moser invariant curves~\cite{Moser56,Silva87,Ozorio89,Harsoula15}.  In this section, we generalize our previous results, remove the dependence on auxiliary structures such as the Moser invariant curves, and derive both exact and approximate formulae to express the periodic orbit action in terms of the homoclinic orbit actions and phase space areas. Detailed numerical verifications are given to demonstrate the accuracy of the procedure. 

\subsection{Exact periodic orbit/homoclinic orbit action differences}
\label{Action formula with entire periodic orbit strings}

Consider an arbitrary unstable periodic orbit $\lbrace y_0 \rbrace \Rightarrow \overline{\gamma}$ for which $n_{\gamma}$ is the primitive period.  The basic idea is to consider the action difference of two auxiliary homoclinic orbits $\lbrace h^{(\gamma\gamma)}_0 \rbrace \Rightarrow \overline{0} \gamma\gamma \overline{0}$ and $\lbrace h^{(\gamma)}_0 \rbrace \Rightarrow \overline{0} \gamma \overline{0}$, and generate its relation to the periodic orbit action ${\cal F}_{\gamma}$. To fix the procedure, split the string $\gamma$ into two substrings: $\gamma = \gamma^{-} \gamma^{+}$. Let the lengths of $\gamma^{-}$ and $\gamma^{+}$ be $n^{-}$ and $n^{+}$, respectively, where $n^{-}+n^{+}=n_{\gamma}$. The splitting can be done arbitrarily, but without loss of generality assume the split is in the middle. If $n_{\gamma}$ is even, $n^{\pm}=\frac{1}{2}n_{\gamma}$; and for cases in which $n_{\gamma}$ is odd, let $n^{-}=n^{+}-1$. Let the zero subscript periodic orbit point be $y_0=y_{n_{\gamma}}\Rightarrow \overline{\gamma^{+}\gamma^{-}}.\overline{\gamma^{+}\gamma^{-}}$, and the zero subscript homoclinic points be
\begin{equation}\label{eq:Homoclinic orbit points symbolic codes association}
\begin{cases}
   & h^{(\gamma)}_0 \Rightarrow \overline{0} \gamma^{-}.\gamma^{+} \overline{0}\\
   & h^{(\gamma\gamma)}_0 \Rightarrow \overline{0} \gamma.\gamma \overline{0} = \overline{0} \gamma^{-}\gamma^{+}.\gamma^{-}\gamma^{+} \overline{0}
  \end{cases}
\end{equation}
from which it follows that 
\begin{equation}
\begin{split}
&h^{(\gamma\gamma)}_{-n^{+}} \Rightarrow \overline{0}\gamma^{-}.\gamma^{+}\gamma^{-}\gamma^{+}\overline{0}\\
&h^{(\gamma\gamma)}_{n^{-}} \Rightarrow \overline{0}\gamma^{-}\gamma^{+}\gamma^{-}.\gamma^{+}\overline{0}
\end{split}
\end{equation}
and $h^{(\gamma)}_{\pm\infty}=h^{(\gamma\gamma)}_{\pm\infty} = x$. With the help of Eq~\eqref{eq:homoclinic action difference2}, the action difference of the two auxiliary homoclinic orbits is given by
\begin{equation}\label{eq:Relative action auxiliary homoclinic orbits entire string}
\begin{split}
&\Delta{\cal F}_{ \overline{0} \gamma\gamma \overline{0},  \overline{0} \gamma \overline{0} }\\
 &=\lim_{N \to \infty} \left[ F\left(h^{(\gamma\gamma)}_{-(N+n^+)},h^{(\gamma\gamma)}_{N+n^-}\right) - F\left(h^{(\gamma)}_{-N},h^{(\gamma)}_{N}\right) \right] \\
 &\quad - n_\gamma {\cal F}_0 \\
&= \lim_{N \to \infty} \left[ F\left(h^{(\gamma\gamma)}_{-(N+n^+)},h^{(\gamma\gamma)}_{-n^+}\right) - F\left(h^{(\gamma)}_{-N},h^{(\gamma)}_0\right) \right]  \\
&\quad + \lim_{N \to \infty}\left[ F\left(h^{(\gamma\gamma)}_{n^-},h^{(\gamma\gamma)}_{N+n^-}\right) - F\left(h^{(\gamma)}_0,h^{(\gamma)}_{N}\right) \right]  \\
&\quad + F\left(h^{(\gamma\gamma)}_{-n^+},h^{(\gamma\gamma)}_{n^-}\right) - n_\gamma {\cal F}_0
\end{split}
\end{equation}
where we have cut $F(h^{(\gamma\gamma)}_{-(N+n^+)},h^{(\gamma\gamma)}_{N+n^-})$ into three parts, $F(h^{(\gamma\gamma)}_{-(N+n^+)},h^{(\gamma\gamma)}_{-n^+})$, $F(h^{(\gamma\gamma)}_{-n^+},h^{(\gamma\gamma)}_{n^-})$ and $F(h^{(\gamma\gamma)}_{n^-},h^{(\gamma\gamma)}_{N+n^-})$, that correspond to the initial, middle, and final parts of $\lbrace h^{(\gamma\gamma)}_0 \rbrace$, respectively. Similarly, $F(h^{(\gamma)}_{-N},h^{(\gamma)}_{N})$ is cut into two parts, $F(h^{(\gamma)}_{-N},h^{(\gamma)}_0)$ and $F(h^{(\gamma)}_0,h^{(\gamma)}_{N})$, corresponding to the initial and final parts of $\lbrace h^{(\gamma)}_0 \rbrace$, respectively.  The choice of the divisions is motivated by the shadowing implied by the similarities of symbolic strings.

The derivation of the action difference of ${\cal F}_{\gamma}$ and $\Delta{\cal F}_{ \overline{0} \gamma\gamma \overline{0},  \overline{0} \gamma \overline{0} }$  proceeds by applying the MacKay-Meiss-Percival action principle, Eq.~\eqref{eq:Meiss92}, separately to the three action difference terms
\begin{equation}
\label{eq:Difference between periodic and homoclinic relative actions entire string}
\begin{split}
&{\cal F}_{\gamma} - \Delta{\cal F}_{ \overline{0} \gamma\gamma \overline{0},  \overline{0} \gamma \overline{0} } \\
&= \lim_{N \to \infty} \left[ F\left(h^{(\gamma)}_{-N},h^{(\gamma)}_0\right) - F\left(h^{(\gamma\gamma)}_{-(N+n^+)},h^{(\gamma\gamma)}_{-n^+}\right) \right]  \\
&\quad + \lim_{N \to \infty}\left[ F\left(h^{(\gamma)}_0,h^{(\gamma)}_{N}\right) -  F\left(h^{(\gamma\gamma)}_{n^-},h^{(\gamma\gamma)}_{N+n^-}\right) \right]  \\
&\quad + \left[ {\cal F}_{\gamma} - F\left(h^{(\gamma\gamma)}_{-n^+},h^{(\gamma\gamma)}_{n^-}\right) \right] + n_\gamma {\cal F}_0
\end{split}
\end{equation}
 and adding their contributions. The end result reduces to a specifiic phase space area.

The first term in Eq.~\eqref{eq:Difference between periodic and homoclinic relative actions entire string} is the difference between the initial parts of the two auxiliary homoclinic orbits.  Let the points $a$ and $b$ of Eq.~\eqref{eq:Meiss92} be $h^{(\gamma\gamma)}_{-(N+n^{+})}$ and $h^{(\gamma)}_{-N}$, respectively, and let the curve $c$ of Eq.~\eqref{eq:Meiss92} be the unstable manifold segment $U[h^{(\gamma\gamma)}_{-(N+n^{+})}\ ,\ h^{(\gamma)}_{-N}]$. After $N$ iterations, $c=U[h^{(\gamma\gamma)}_{-(N+n^{+})}\ ,\ h^{(\gamma)}_{-N}]$ is mapped to $c^{\prime} = U[h^{(\gamma\gamma)}_{-n^{+}}\ ,\ h^{(\gamma)}_{0}]$, and this leads to
\begin{equation}\label{eq:Homoclinic decomposition single homoclinic first segment}
\begin{split}
& \lim_{N \to \infty}\left [ F\left(h^{(\gamma)}_{-N}\ ,\ h^{(\gamma)}_{0}\right)  - F\left(h^{(\gamma\gamma)}_{-(N+n^{+})}\ ,\ h^{(\gamma\gamma)}_{-n^{+}} \right) \right] \\
&=\int\limits_{U[h^{(\gamma\gamma)}_{-n^{+}},h^{(\gamma)}_{0}]}p\mathrm{d}q- \lim_{N \to \infty}\int\limits_{U[h^{(\gamma\gamma)}_{-(N+n^{+})}, h^{(\gamma)}_{-N}]}p\mathrm{d}q \\
& = \int\limits_{U[h^{(\gamma\gamma)}_{-n^{+}},h^{(\gamma)}_{0}]}p\mathrm{d}q \ .
\end{split}
\end{equation}
Similarly, the second term is the action difference between the final parts of the two auxiliary homoclinic orbits. By the same logic with $a=h^{(\gamma\gamma)}_{n^{-}}$\ , $b=h^{(\gamma)}_0$, and $c=S[h^{(\gamma\gamma)}_{n^{-}}\ ,\ h^{(\gamma)}_0]$,
\begin{equation}
\label{eq:Homoclinic decomposition single homoclinic third segment}
\begin{split}
& \lim_{N \to \infty} \left[F\left(h^{(\gamma)}_0\ ,\ h^{(\gamma)}_N\right) - F\left(h^{(\gamma\gamma)}_{n^{-}}\ ,\ h^{(\gamma\gamma)}_{N+n^{-}}\right) \right]\\
&= \int\limits_{S[h^{(\gamma)}_0\ ,\ h^{(\gamma\gamma)}_{n^{-}}]}p\mathrm{d}q\ .
\end{split}
\end{equation}
The third term in Eq.~\eqref{eq:Difference between periodic and homoclinic relative actions entire string} is the difference between the periodic orbit action ${\cal F}_\gamma=F(y_0,y_{n_{\gamma}})$, and the middle $n_\gamma$ iterations of $\lbrace h^{(\gamma\gamma)}_0 \rbrace$, $F\left(h^{(\gamma\gamma)}_{-n^{+}}\ ,\ h^{(\gamma\gamma)}_{n^{-}} \right)$. Choose the points $a$ and $b$ of Eq.~\eqref{eq:Meiss92} to be $h^{(\gamma\gamma)}_{-n^{+}}$ and $y_{0}$, respectively, and choose the curve $c$ to be an arbitrary curve $C[h^{(\gamma\gamma)}_{-n^{+}},y_0]$ connecting them. As $y_0$ is a fixed point under $n_\gamma$ iterations, $c$ maps to $c^{\prime}=C^{\prime}[h^{(\gamma\gamma)}_{n^{-}}\ ,\ y_0]$. Thus,
\begin{equation}
\label{eq:Homoclinic decomposition single homoclinic second segment}
\begin{split}
&{\cal F}_\gamma - F\left(h^{(\gamma\gamma)}_{-n^{+}}\ ,\ h^{(\gamma\gamma)}_{n^{-}} \right)\\
&=\int\limits_{C^{\prime}[h^{(\gamma\gamma)}_{n^{-}}\ ,\ y_0]}p\mathrm{d}q-\int\limits_{C[h^{(\gamma\gamma)}_{-n^{+}}\ ,\ y_{0}]}p\mathrm{d}q\\
&= \int\limits_{C[y_{0}\ ,\ h^{(\gamma\gamma)}_{-n^{+}}]}p\mathrm{d}q + \int\limits_{C^{\prime}[h^{(\gamma\gamma)}_{n^{-}}\ ,\ y_0]}p\mathrm{d}q \ .
\end{split}
\end{equation}
Although the curve $C[h^{(\gamma\gamma)}_{-n^{+}}\ ,\ y_{0}]$ can be an arbitrary curve connecting the end points, a convenient choice without loss of generality is to let $C$ be the straight line segment as shown in Fig.~\ref{fig:Area_correction_single}.  

From the discussion in Appendix~\ref{Markov partition}, it is clear that $h^{(\gamma\gamma)}_{-n^{+}} \Rightarrow \overline{0}\gamma^{-}.\gamma^{+}\gamma^{-}\gamma^{+}\overline{0}$ is within an exponentially small ($\sim O(e^{-\mu n_{\gamma} })$) neighborhood of $y_0$ and $C$ is exponentially close to the stable direction of $y_0$.  The images of $h^{(\gamma\gamma)}_{-n^{+}}$ and $y_0$ under successive forward iterations first approach and then separate from each other.  After $n_{\gamma}$ iterations, the final image $h^{(\gamma\gamma)}_{n^{-}} \Rightarrow \overline{0}\gamma^{-}\gamma^{+}\gamma^{-}.\gamma^{+}\overline{0}$ is exponentially close to the unstable direction of $y_0$ and remains within a small neighborhood ($\sim O(e^{-\mu  n_{\gamma} })$) of it.  Since under the local linearized map of $M^{n_{\gamma}}$, the straight line segment $C$ must be mapped into another straight line segment, the image $C^{\prime}=M^{n_{\gamma}}(C)$ must be nearly a straight line as well.  The geometry is shown in Fig.~\ref{fig:Area_correction_single}. 
\begin{figure}[ht]
\centering
{\includegraphics[width=5cm]{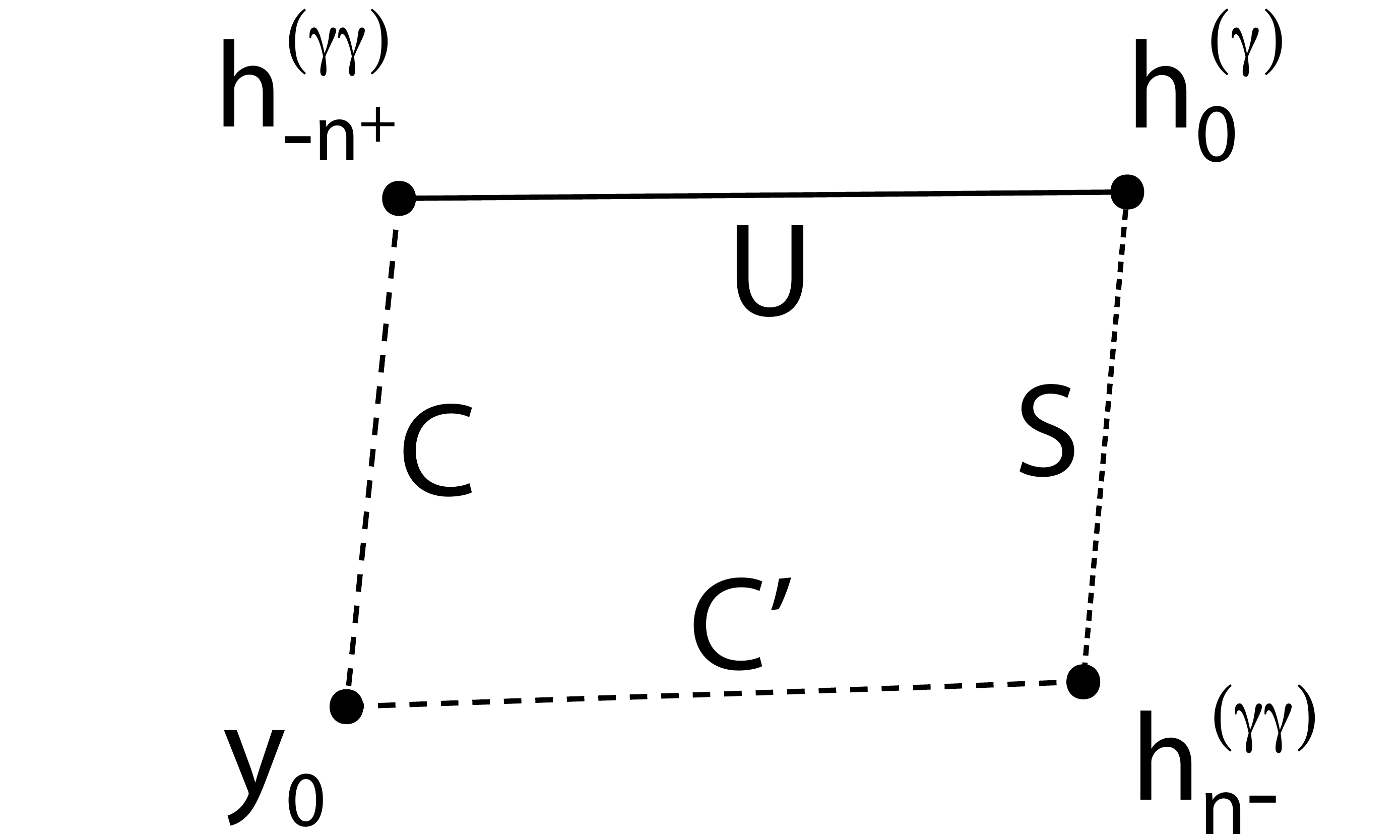}}
 \caption{(Schematic) The correction term in Eq.~\eqref{eq:Homoclinic decomposition action formula single exact} is determined by this exponentially small near-parallelogram area. The lower-left corner $y_0=y_{n_{\gamma}} \Rightarrow \overline{\gamma^{+}\gamma^{-}}.\overline{\gamma^{+}\gamma^{-}}$ is a fixed point under $M^{n_{\gamma}}$. The other three corners are homoclinic points $h^{(\gamma\gamma)}_{-n^{+}} \Rightarrow \overline{0} \gamma^{-}.\gamma^{+}\gamma^{-}\gamma^{+} \overline{0}$\ ,\ $h^{(\gamma)}_0 \Rightarrow \overline{0} \gamma^{-}.\gamma^{+} \overline{0}$ and $h^{(\gamma\gamma)}_{n^{-}} \Rightarrow \overline{0} \gamma^{-}\gamma^{+}\gamma^{-}.\gamma^{+} \overline{0}$\ . The unstable and stable segments between corresponding homoclinic points are labeled by $U$ and $S$, respectively. $C$ is a straight line segment connecting $h^{(\gamma\gamma)}_{-n^{+}}$ and $y_0$, which is exponentially close to the stable direction of $y_0$ .  Under $M^{n_{\gamma}}$,  $C$ is mapped into a near-straight segment $C^{\prime}$ connecting $h^{(\gamma\gamma)}_{n^{-}}$ and $y_{n_{\gamma}}=y_0$, which is exponentially close to the unstable direction of $y_0$.  Under $n_{\gamma}$ iterations, the successive images of $y_0$ and $h^{(\gamma\gamma)}_{-n^{+}} $ first approach, then separate from each other, making a near fly-by somewhere in the middle. }
\label{fig:Area_correction_single}
\end{figure}  

Adding the results of Eqs.~\eqref{eq:Homoclinic decomposition single homoclinic first segment} - \eqref{eq:Homoclinic decomposition single homoclinic second segment}, substituting into Eq.~\eqref{eq:Difference between periodic and homoclinic relative actions entire string} and rearranging terms gives an exact formula expressing the periodic orbit action in terms of the relative homoclinic orbit actions, a multiple of the fixed point's action, plus a phase space area: 
\begin{equation}
\label{eq:Homoclinic decomposition action formula single exact}
{\cal F}_{\gamma} = n_\gamma {\cal F}_0 + {\Delta}{\cal F}_{ \overline{0} \gamma\gamma \overline{0} ,  \overline{0} \gamma \overline{0} } + {\cal A}^\circ_{CUSC^{\prime}[y_0 , h^{(\gamma\gamma)}_{-n^{+}} , h^{(\gamma)}_0 , h^{(\gamma\gamma)}_{n^{-}}]}
\end{equation}
where
\begin{equation}
\begin{split}
&{\cal A}^\circ_{CUSC^{\prime}[y_0\ ,\  h^{(\gamma\gamma)}_{-n^{+}}\ ,\  h^{(\gamma)}_0\ ,\  h^{(\gamma\gamma)}_{n^{-}}]}=\\
&\int\limits_{C[y_{0}\ ,\ h^{(\gamma\gamma)}_{-n^{+}}]}p\mathrm{d}q+\int\limits_{U[h^{(\gamma\gamma)}_{-n^{+}}\ ,\ h^{(\gamma)}_{0}]}p\mathrm{d}q \\
&+  \int\limits_{S[h^{(\gamma)}_0\ ,\ h^{(\gamma\gamma)}_{n^{-}}]}p\mathrm{d}q+\int\limits_{C^{\prime}[h^{(\gamma\gamma)}_{n^{-}}\ ,\ y_0]}p\mathrm{d}q
\end{split}
\end{equation}
yields the area of the near-parallelogram in Fig.~\ref{fig:Area_correction_single}.  This result is invariant under all possible ways of partitioning $\gamma$ into $\gamma=\gamma^{-}\gamma^{+}$, as long as the $y_0$, $h^{(\gamma)}_0$ and $h^{(\gamma\gamma)}_0$ are defined consistently. Nevertheless, the choice made in the derivation, where $\gamma^{-}$ and $\gamma^{+}$ have near-identical lengths, leads to a near-parallelogram shaped region in Fig.~\ref{fig:Area_correction_single}. For other choices, especially those with greatly unequal lengths, $\gamma^{-}$ and $\gamma^{+}$, the ${\cal A}^\circ_{CUSC^{\prime}[y_0\ ,\  h^{(\gamma\gamma)}_{-n^{+}}\ ,\  h^{(\gamma)}_0\ ,\  h^{(\gamma\gamma)}_{n^{-}}]}$ area would be distorted into long thin, possibly strongly curved, strips, making it more difficult to calculate the area integral. In addition, it is possible to give an order estimate for the integral, which is best done by choosing $\gamma^{-}$ and $\gamma^{+}$ to have the same lengths. 

A careful inspection on the symbolic codes of the four corners of the parallelogram allows us to estimate the order of magnitude of its area. Notice the four corner have symbolic codes $y_0 \Rightarrow \overline{\gamma^{+}\gamma^{-}}.\overline{\gamma^{+}\gamma^{-}}$\ , $h^{(\gamma\gamma)}_{-n^{+}} \Rightarrow \overline{0}\gamma^{-}.\gamma^{+}\gamma^{-}\gamma^{+}\overline{0}$\ , $h^{(\gamma)}_0 \Rightarrow \overline{0}\gamma^{-}.\gamma^{+}\overline{0}$\ , and $h^{(\gamma\gamma)}_{n^{-}} \Rightarrow \overline{0}\gamma^{-}\gamma^{+}\gamma^{-}.\gamma^{+}\overline{0}$. Since $\gamma^{-}$ and $\gamma^{+}$ are chosen to have near-identical lengths, the symbolic codes of the four corners will match along central block lengths of at least $n_{\gamma}$. Therefore, Eq.~\eqref{eq:Matching central block lengths area estimate} implies an upper bound for the order of its magnitude:
\begin{equation}\label{eq:Homoclinic decomposition action formula single area correction estimate}
{\cal A}^\circ_{CUSC^{\prime}[y_0 , h^{(\gamma\gamma)}_{-n^{+}} , h^{(\gamma)}_0 , h^{(\gamma\gamma)}_{n^{-}}]} \sim O(e^{-\mu n_{\gamma}})
\end{equation}
where $\mu$ is the Lyapunov exponent of the system. For long orbits ($n_\gamma$ large enough), this area can be neglected and leads to an approximate form of Eq.~\eqref{eq:Homoclinic decomposition action formula single exact}:
\begin{eqnarray}
\label{eq:Homoclinic decomposition action formula single approximate}
{\cal F}_{\gamma} &=& n_{\gamma}{\cal F}_0 + {\Delta}{\cal F}_{ \overline{0} \gamma\gamma \overline{0} ,  \overline{0} \gamma \overline{0} } + O(e^{-\mu n_{\gamma}}) \nonumber \\
&=& n_{\gamma}{\cal F}_0 + {\cal A}^\circ_{US[h^{(\gamma)}_0 , h^{(\gamma\gamma)}_0 ]} + O(e^{-\mu n_{\gamma}}) 
\end{eqnarray}
that expresses the periodic orbit action in terms of the fixed point action and the relative action between corresponding homoclinic orbits with well-estimated error.  Notice that the right hand side of Eq.~(\ref{eq:Homoclinic decomposition action formula single approximate}) is evaluated without constructing the periodic orbit or locating its phase-space points.  In~\cite{Li17} a method was given to calculate the homoclinic orbits in very stable way.  The basic idea is that calculating successive intersections of the stable and unstable manifolds is a structurally stable operation whereas direct propagation of phase points magnifies small propagation errors exponentially fast.  Thus, Eq.~\eqref{eq:Homoclinic decomposition action formula single approximate} can be used to calculate the classical action of arbitrarily long periodic orbits. The periodic orbit approximated is the one which follows the excursion of $\lbrace h^{(\gamma)}_0 \rbrace$. 

\subsection*{Optimal representation and numerical verification }
\label{Optimal representation and numerical verification}

Equation~\eqref{eq:Homoclinic decomposition action formula single area correction estimate} gives an upper bound of the area correction order of magnitude corresponding to the worst case scenario, which results if the string $\gamma$ ends with the digit ``$1$'' on the left and right sides.  However, if $\gamma$ has $L_{\gamma}$ consecutive ``$0$"s on its left end, and $R_{\gamma}$ consecutive ``$0$"s on its right end, the lengths of matching central blocks of the symbolic codes of the four corners increases to $(n_{\gamma}+L_{\gamma}+R_{\gamma})$ and the error estimate of Eq.~(\ref{eq:Homoclinic decomposition action formula single area correction estimate}) improves to
\begin{equation}\label{eq:Homoclinic decomposition action formula single area correction estimate improved}
{\cal A}^\circ_{CUSC^{\prime}[y_0 , h^{(\gamma\gamma)}_{-n^{+}} , h^{(\gamma)}_0 , h^{(\gamma\gamma)}_{n^{-}}]} \sim O(e^{-\mu ( n_{\gamma} + L_{\gamma} + R_{\gamma} )})\ .
\end{equation}
Therefore, given a periodic orbit $\lbrace y \rbrace \Rightarrow \overline{\gamma}$ with length $n_{\gamma}$, not all cyclic permutations of $\gamma$ necessarily lead to the same quaility approximation in Eq.~(\ref{eq:Homoclinic decomposition action formula single approximate}) due to  Eq.~\eqref{eq:Homoclinic decomposition action formula single area correction estimate improved}. 

The optimal situation is to cyclically permute $\gamma$ until it has the longest string of ``0''s on the left or right boundary (or some combination). Let us consider a concrete example for the H\'enon map with $\lbrace y_0 \rbrace \Rightarrow \overline{10001}$, $\gamma=10001$ (worst case scenario), and an optimal scenario $\gamma^{\prime}=00011$.  They give rise to different homoclinic orbits in Eq.~\eqref{eq:Homoclinic decomposition action formula single approximate} since
\begin{equation}\label{eq:Multiple accompanying homoclinic orbits}
\begin{split}
&\overline{0} \gamma \overline{0}  =  \overline{0}10001\overline{0} \\
&\overline{0} \gamma^{\prime} \overline{0} =  \overline{0}11\overline{0} \ 
\end{split}
\end{equation}
which do lead to differing quality of approximation. 

The $\overline{10001}$ orbit, starts from point $(3.18110104534044, 3.18110104534044)$ and maps back into itself to 12 decimal places in a double precision calculation after $5$ iterations. Direct calculation of the classical action function gives:
\begin{equation}\label{eq:satellite orbit action general approx symbolic codes example exact}
{\cal F}_{10001}=34.093709790589912\ . 
\end{equation} 
Of course, for the exact expression there is no difference in which homoclinic orbits which are used.  We first verify Eq.~\eqref{eq:Homoclinic decomposition action formula single exact} with $\gamma^{\prime}=00011$, and partition it into $\gamma^{\prime -}=00$ and $\gamma^{\prime +}=011$.  Substituting the three symbols into Eq.~\eqref{eq:Homoclinic decomposition action formula single exact} leads to
\begin{equation}\label{eq:satellite orbit action general exact symbolic codes example}
\begin{split}
&{\cal F}_{10001} = 5{\cal F}_{0} + \Delta {\cal F}_{ \overline{0} 1100011 \overline{0} , \overline{0} 11 \overline{0}}\\
&+ {\cal A}^\circ_{CUSC^{\prime}[y^{\prime}_0 , h^{(\gamma^{\prime}\gamma^{\prime})}_{-n^{+}} , h^{(\gamma^{\prime})}_0 , h^{(\gamma^{\prime}\gamma^{\prime})}_{n^{-}}]}\\
&=34.093709790630861\ . 
\end{split}
\end{equation}
The relative error is one part in $10^{-12}$, which is as good as possible given the propagation error.  This ``exact'' calculation required finding the point $y^{\prime}_0$ corresponding to $\overline{01100}.\overline{01100}$, an impractical task for very long orbits.

Next, consider Eq.~(\ref{eq:Homoclinic decomposition action formula single approximate}) for both the non-optimal representation $\gamma=10001$ and the optimal representation $\gamma^{\prime}=00011$.  There is the non-optimal approximation
\begin{equation}
\label{eq:satellite orbit action general approx symbolic codes example approx 1}
\begin{split}
{\cal F}_{10001}&\approx 5{\cal F}_{ 0 } + \Delta {\cal F}_{ \overline{0} 1000110001 \overline{0} , \overline{0} 10001 \overline{0} }\\
&=34.091429013921982 
\end{split}
\end{equation}
and the optimal
\begin{equation}
\label{eq:satellite orbit action general approx symbolic codes example approx 2}
\begin{split}
{\cal F}_{10001}&\approx 5{\cal F}_{ 0 } + \Delta {\cal F}_{ \overline{0} 1100011 \overline{0} , \overline{0} 11 \overline{0} }\\
&=34.093701415127327 \ .
\end{split}
\end{equation}
The non-optimal case leads to a relative error of $6.7 \times 10^{-5}$, whereas the optimal case gives 
a relative error of $2.5 \times 10^{-7}$, a factor $270$ smaller error.

The choice of optimal representation is not necessarily unique as occurs if $\gamma$ has multiple substrings with the same maximum possible number of consecutive ``$0$"s.  For example, let $ \overline{\gamma} = \overline{110011001} $, which has two ``$00$" substrings. Following the above procedure, we can identify two optimal representations, namely $\gamma_1=001100111$ and $\gamma_2=001110011$, respectively.  In these cases, the two optimal accompanying orbits should yield errors within the same order of magnitude, thus equally valid in practice. 

\subsection{Action formulae with partitioned substrings}
\label{Action formula with reshuffled substrings}

In Sec.~\ref{Action formula with entire periodic orbit strings}, the approximate action formula, Eq.~\eqref{eq:Homoclinic decomposition action formula single approximate}, requires either the construction of the homoclinic orbit $ \overline{0} \gamma \gamma \overline{0}$, or the area integral involving the homoclinic point $h^{(\gamma\gamma)}_0 \Rightarrow \overline{0} \gamma . \gamma \overline{0}$ as one of the end points.  Although these quantities can be calculated in quite stable ways without exponentially diverging error, the calculation is time-consuming for orbits with large periods.  Upon further investigation into the geometric relations between classical orbit structures, it turns out to be possible to partition the symbolic codes, thereby relying on homoclinic orbits with shorter excursions, the longest of which are half the length of those in Sect.~\ref{Action formula with entire periodic orbit strings}.  This generates a host of new relations amongst orbits and reduces the complexity of the task. 

Starting from a long periodic orbit $\lbrace y \rbrace \Rightarrow \overline{\gamma} $, cut its symbolic string into two substrings in: $\gamma = \gamma_1 \gamma_2$, such that their lengths $n_{\gamma}=n_1+n_2$, where $n_1$ and $n_2$ are the lengths of $\gamma_1$ and $\gamma_2$, respectively. Assuming neither $\gamma_1$ nor $\gamma_2$ is a single digit string, further cut both of them into two substrings, such that $\gamma_1=\gamma^{-}_1\gamma^{+}_1$ and $\gamma_2=\gamma^{-}_2\gamma^{+}_2$.  Let the length of $\gamma^{\pm}_{i}$ be $n^{\pm}_{i}$, then $n_i=n^{-}_i+n^{+}_i$ ($i=1,2$).  For convenience, assume cutting the $\gamma_i$s ($i=1,2$) in the middle, similar to Sect.~\ref{Action formula with entire periodic orbit strings}.  Then the periodic orbit can be denoted alternatively by $\lbrace y \rbrace \Rightarrow  \overline{ \gamma^{-}_1\gamma^{+}_1 \gamma^{-}_2\gamma^{+}_2 } $.  To fix $y_0$, position the decimal point in the string $\gamma^{-}_1.\gamma^{+}_1 \gamma^{-}_2\gamma^{+}_2 $, and the other $y_n$ follow by the appropriate shifts in the decimal point.

Four homoclinic orbits are involved in the determination of the classical action ${\cal F}_{\gamma}$. These auxiliary homoclinic orbits are
\begin{equation}\label{eq:Accompanying homoclinic orbits reshuffled substrings}
\begin{split}
   & \lbrace h^{(\gamma_1)}_0 \rbrace \Rightarrow  \overline{0} \gamma_1 \overline{0} \\
   & \lbrace h^{(\gamma_2)}_0 \rbrace \Rightarrow  \overline{0} \gamma_2 \overline{0} \\
   & \lbrace h^{(\gamma_1 \gamma_2)}_0 \rbrace \Rightarrow \overline{0} \gamma_1 \gamma_2 \overline{0} \\
   & \lbrace h^{(\gamma_2 \gamma_1)}_0 \rbrace \Rightarrow \overline{0} \gamma_2 \gamma_1 \overline{0} 
  \end{split}
\end{equation}
where the symbolic codes of $h^{(\gamma_1)}_0$\ , $h^{(\gamma_2)}_0$\ , $h^{(\gamma_1 \gamma_2)}_0$ and $h^{(\gamma_2 \gamma_1)}_0$ are identified in Eq.~\eqref{eq:Acompanying homoclinic orbit points setting appendix}.

As shown in Appendix.~\ref{Derivation of Equation}, following similar but generalized steps from Sec.~\ref{Action formula with entire periodic orbit strings} gives a generalized exact analytic formula for the periodic orbit action:
\begin{equation}\label{eq:Homoclinic decomposition action formula reshuffle exact}
\begin{split}
{\cal F}_{\gamma} &= n_{\gamma}{\cal F}_{0} + {\Delta}{\cal F}_{ \overline{0} \gamma_1\gamma_2 \overline{0} ,  \overline{0} \gamma_1 \overline{0} }  +  {\Delta}{\cal F}_{ \overline{0} \gamma_2\gamma_1 \overline{0} ,  \overline{0} \gamma_2 \overline{0} }\\
& + {\cal A}^\circ_{ C U S C^{\prime} [ y_0\ ,\ h^{ (\gamma_1\gamma_2) }_{ -n^{+}_1 }\ ,\ h^{ (\gamma_1) }_0\ ,\ h^{ (\gamma_2\gamma_1) }_{ n^{-}_1 } ] } \\
& + {\cal A}^\circ_{ C U S C^{\prime} [ y_{ n^{+}_1 + n^{-}_2 }\ ,\ h^{ (\gamma_2\gamma_1) }_{ -n^{+}_2 }\ ,\ h^{ (\gamma_2) }_0\ ,\ h^{ (\gamma_1\gamma_2) }_{ n^{-}_2 } ] } 
 \end{split}
\end{equation}
where the two near-parallelogram area terms ${\cal A}^\circ_{ C U S C^{\prime} [\cdots] }$ have similar interpretations as in Eq.~\eqref{eq:Homoclinic decomposition action formula single exact}. Notice the symbolic codes of the four corners of the first and the second parallelogram match in central block lengths of at least $n_1$ and $n_2$, respectively. Therefore, we can estimate the upper bounds of the orders of magnitudes of these areas as
\begin{equation}\label{eq:Homoclinic decomposition action formula reshuffle area corrections estimate 1}
{\cal A}^\circ_{ C U S C^{\prime} [ y_0 , h^{ (\gamma_1\gamma_2) }_{ -n^{+}_1 } , h^{ (\gamma_1) }_0 , h^{ (\gamma_2\gamma_1) }_{ n^{-}_1 } ] } \sim O( e^{ -\mu n_1 } ) 
\end{equation}
and
\begin{equation}\label{eq:Homoclinic decomposition action formula reshuffle area corrections estimate 2}
{\cal A}^\circ_{ C U S C^{\prime} [ y_{ n^{+}_1 + n^{-}_2 } , h^{ (\gamma_2\gamma_1) }_{ -n^{+}_2 } , h^{ (\gamma_2) }_0 , h^{ (\gamma_1\gamma_2) }_{ n^{-}_2 } ] } \sim O( e^{ -\mu n_2 } ) 
\end{equation}
which lead to the approximation:
 \begin{equation}\label{eq:Homoclinic decomposition action formula reshuffle approximate}
\begin{split}
{\cal F}_{\gamma} &= n_{\gamma}{\cal F}_{0} + {\Delta}{\cal F}_{ \overline{0} \gamma_1\gamma_2 \overline{0} ,  \overline{0} \gamma_1 \overline{0} }  +  {\Delta}{\cal F}_{ \overline{0} \gamma_2\gamma_1 \overline{0} ,  \overline{0} \gamma_2 \overline{0} }\\
& + O( e^{ -\mu n_1 }  +  e^{ -\mu n_2 } ) 
 \end{split}
\end{equation}
which expresses ${\cal F}_{\gamma}$ in terms of two relative homoclinic orbit actions, but of shorter excursions.  As before, a geometric alternative exists that does not require the calculation of the homoclinic orbits. Using Eq.~\eqref{eq:homoclinic action difference} results in
\begin{equation}
\label{eq:Homoclinic decomposition reshuffle area action relation}
\begin{split}
&{\Delta}{\cal F}_{ \overline{0} \gamma_1\gamma_2 \overline{0} ,  \overline{0} \gamma_1 \overline{0} }  +  {\Delta}{\cal F}_{ \overline{0} \gamma_2\gamma_1 \overline{0} ,  \overline{0} \gamma_2 \overline{0} }\\
&={\cal A}^\circ_{US[ h^{(\gamma_1)}_0 , h^{(\gamma_1\gamma_2)}_0 ]} + {\cal A}^\circ_{US[ h^{(\gamma_2)}_0 , h^{(\gamma_2\gamma_1)}_0 ]}\\
&= \int\limits_{U[ h^{ ( \gamma_1 ) }_{ 0 },h^{ ( \gamma_1\gamma_2 ) }_{0} ]} p\mathrm{d}q +  \int\limits_{S[ h^{ ( \gamma_1\gamma_2 ) }_{ 0 },h^{ ( \gamma_1 ) }_{0} ]} p\mathrm{d}q\\
&\quad + \int\limits_{U[ h^{ ( \gamma_2 ) }_{ 0 },h^{ ( \gamma_2\gamma_1 ) }_{0} ]} p\mathrm{d}q +  \int\limits_{S[ h^{ ( \gamma_2\gamma_1 ) }_{ 0 },h^{ ( \gamma_2 ) }_{0} ]} p\mathrm{d}q
\end{split}
\end{equation}
and a further manipulation of the integral paths combines the two areas into a single curvy-parallelogram. This is done by splitting the two stable manifolds paths
\begin{equation}\label{eq:Splitting the stable manifold path 1}
\int\limits_{S[ h^{ ( \gamma_1\gamma_2 ) }_{ 0 },h^{ ( \gamma_1 ) }_{0} ]} p\mathrm{d}q= \int\limits_{S[ h^{ ( \gamma_1\gamma_2 ) }_{ 0 },h^{ ( \gamma_2 ) }_{0} ]} p\mathrm{d}q + \int\limits_{S[ h^{ ( \gamma_2 ) }_{ 0 },h^{ ( \gamma_1 ) }_{0} ]} p\mathrm{d}q
\end{equation}
and 
\begin{equation}\label{eq:Splitting the stable manifold path 2}
\int\limits_{S[ h^{ ( \gamma_2\gamma_1 ) }_{ 0 },h^{ ( \gamma_2 ) }_{0} ]} p\mathrm{d}q= \int\limits_{S[ h^{ ( \gamma_2\gamma_1 ) }_{ 0 },h^{ ( \gamma_1 ) }_{0} ]} p\mathrm{d}q + \int\limits_{S[ h^{ ( \gamma_1 ) }_{ 0 },h^{ ( \gamma_2 ) }_{0} ]} p\mathrm{d}q\ .
\end{equation}
Substituting Eqs.~\eqref{eq:Splitting the stable manifold path 1} and \eqref{eq:Splitting the stable manifold path 2} into Eq.~\eqref{eq:Homoclinic decomposition reshuffle area action relation}, and notice cancellations between certain paths, we obtain
\begin{equation}\label{eq:Area action principle for four homoclinic points}
\begin{split}
&{\Delta}{\cal F}_{ \overline{0} \gamma_1\gamma_2 \overline{0} ,  \overline{0} \gamma_1 \overline{0} }  +  {\Delta}{\cal F}_{ \overline{0} \gamma_2\gamma_1 \overline{0} ,  \overline{0} \gamma_2 \overline{0} }\\
&={\cal A}^\circ_{US[ h^{(\gamma_1)}_0 , h^{(\gamma_1\gamma_2)}_0 ]} + {\cal A}^\circ_{US[ h^{(\gamma_2)}_0 , h^{(\gamma_2\gamma_1)}_0 ]}\\
&={\cal A}^\circ_{SUSU[ h^{(\gamma_1\gamma_2)}_0 , h^{(\gamma_2)}_0 , h^{(\gamma_2\gamma_1)}_0 , h^{(\gamma_1)}_0 ]}
\end{split}
\end{equation}
and therefore a geometric alternative of Eq.~\eqref{eq:Homoclinic decomposition action formula reshuffle approximate}
\begin{equation}\label{eq:Homoclinic decomposition action formula reshuffle approximate geometric}
\begin{split}
{\cal F}_{\gamma} &= n_{\gamma}{\cal F}_{0} + {\cal A}^\circ_{SUSU[ h^{(\gamma_1\gamma_2)}_0 , h^{(\gamma_2)}_0 , h^{(\gamma_2\gamma_1)}_0 , h^{(\gamma_1)}_0 ]}\\
& + O( e^{ -\mu n_1 } + e^{ -\mu n_2 } )
\end{split}
\end{equation}
which only requires integration along the corresponding stable and unstable manifold segments. Notice that Eq.~\eqref{eq:Area action principle for four homoclinic points} is actually the classical result described in Eq.~(8.16) of Ref.~\cite{Meiss92}, which expresses the action relations between two pairs of homoclinic orbits in terms of the region bounded by alternating stable and unstable manifolds connecting them. 

For completeness, we mention that it is possible to partition $\gamma$ into three or more substrings.  For example, the triple partition, $\gamma=\gamma_1\gamma_2\gamma_3$, leads to:
\begin{equation}\label{eq:Homoclinic decomposition action formula reshuffle approximate 3 substrings}
\begin{split}
&{\cal F}_{ \gamma} = n_{\gamma}{\cal F}_{0} + {\Delta}{\cal F}_{ \overline{0}\gamma_1\gamma_2\gamma_3\overline{0} , \overline{0}\gamma_1\gamma_2\overline{0} } \\
& + {\Delta}{\cal F}_{ \overline{0}\gamma_2\gamma_3\gamma_1\overline{0} , \overline{0}\gamma_2\gamma_3\overline{0} }+{\Delta}{\cal F}_{ \overline{0}\gamma_3\gamma_1\gamma_2\overline{0} , \overline{0}\gamma_3\gamma_1\overline{0} }\\
& + O( e^{ -\mu (n_1+n_2) } + e^{ -\mu (n_2+n_3) } + e^{ -\mu (n_3+n_1) } ) \ .
 \end{split}
\end{equation}
Note that this formula also has an exact version similar to Eq.~\eqref{eq:Homoclinic decomposition action formula reshuffle exact} consisting of three ${\cal A}^\circ_{ C U S C^{\prime} [\cdots] }$ correction terms.  The derivation follows by a straightforward generalization of the procedure in Appendix~\ref{Derivation of Equation}.  Similarly, in the $M$-tuple partition case, $\gamma=\gamma_1\gamma_2\cdots\gamma_M$, the approximation is:
 \begin{equation}
 \label{eq:Homoclinic decomposition action formula reshuffle approximate M substrings}
\begin{split}
{\cal F}_{ \gamma} =& n_{\gamma}{\cal F}_{0}  + \sum_{(i_1 \cdots i_M)}  {\Delta}{\cal F}_{ \overline{0}\gamma_{i_1} \cdots \gamma_{i_M}\overline{0}\ ,\  \overline{0}\gamma_{i_1} \cdots \gamma_{i_{M-1}}\overline{0}} \\
& + O \left( \sum_{(i_1 \cdots i_M)} e^{ -\mu (n_{i_1} + \cdots +n_{i_{M-1}} ) } \right)
 \end{split}
\end{equation}
where $\sum_{(i_1 \cdots i_M)}$ denotes the sum over all cyclic permutations of $(1,2,\cdots,M)$, therefore consisting of $M$ terms.

\subsection*{Optimal partition and numerical verification}    
\label{Optimal partition and numerical verification}

It is of interest to know, given a periodic orbit $\overline{\gamma}$, which among all possible ways of partitioning $\gamma$ into $\gamma_1\gamma_2$ leads to an optimal partition for the approximation, Eq.~\eqref{eq:Homoclinic decomposition action formula reshuffle approximate}. 

Similar to Sec.~\ref{Optimal representation and numerical verification}, the key is the order of magnitudes of the two ${\cal A}^\circ_{CUSC^{\prime}[\cdots]}$ areas. Just as for Eq.~(\ref{eq:Homoclinic decomposition action formula single area correction estimate}), Eqs.~\eqref{eq:Homoclinic decomposition action formula reshuffle area corrections estimate 1} and \eqref{eq:Homoclinic decomposition action formula reshuffle area corrections estimate 2} give the upper bounds of the worst case scenarios where $\gamma_2$ and $\gamma_1$ have both digits ``1" on their left and right ends. For other cases where there are $0$s on either the left or right ends of $\gamma_2$ or $\gamma_1$, the estimates can be further improved to:
\begin{equation}
\label{eq:Homoclinic decomposition action formula reshuffle area corrections estimate elaborated}
\begin{split}
& {\cal A}^\circ_{ C U S C^{\prime} [ y_0 , h^{ (\gamma_1\gamma_2) }_{ -n^{+}_1 } , h^{ (\gamma_1) }_0 , h^{ (\gamma_2\gamma_1) }_{ n^{-}_1 } ] } \sim O( e^{ -\mu ( n_1 + L_2 + R_2 ) } )\ ,\\
&  {\cal A}^\circ_{ C U S C^{\prime} [ y_{ (n^{+}_1 + n^{-}_2 ) } , h^{ (\gamma_2\gamma_1) }_{ -n^{+}_2 } , h^{ (\gamma_2) }_0 , h^{ (\gamma_1\gamma_2) }_{ n^{-}_2 } ] }  \sim O( e^{ -\mu ( n_2 + L_1 + R_1 ) } ) 
\end{split}
\end{equation} 
where $L_i$ and $R_i$ are the total numbers of consecutive ``$0$"s counted starting from the very left and right ends, respectively, on the substring $\gamma_i$ ($i=1,2$).  The error associated with Eq.~\eqref{eq:Homoclinic decomposition action formula reshuffle approximate} is determined by the lesser of the two exponents. Thus, of all partitions $\gamma^{\prime}=\gamma^{\prime}_1\gamma^{\prime}_2$, the one which maximizes the $\min(n_1+L_2+R_2\ ,\ n_2+L_1+R_1)$ yields the smallest error. 

For a numerical verification, we have calculated the period-$8$ orbit $\lbrace y \rbrace \Rightarrow \overline{\gamma} = \overline{10110000} $, for which $y_0=(2.9268794696022995, -1.7889675999506438)$.  The action function calculated following the orbit is
\begin{equation}
\label{eq:Homoclinic decomposition action formula reshuffle numerical periodic orbit action}
{\cal F}_{10110000}=50.526431207035948\ .
\end{equation} 

A non-optimal partition of this orbit is chosen to be $\gamma_1=1011$ and $\gamma_2=0000$.  For this partition $\min(4+4,4+0)=4$.  Note that $\gamma_2=0000$ actually corresponds to the fixed-point being iterated four times.  An optimal partition is $\gamma^{\prime}_1=1100$ and $\gamma^{\prime}_2=0010$, for which $\min(4+3,4+2)=6$.  Evaluating both partitions with Eq.~\eqref{eq:Homoclinic decomposition action formula reshuffle approximate} gives, the non-optimal result,
\begin{equation}\label{eq:Homoclinic decomposition action formula reshuffle numerical non optimal}
\begin{split}
{\cal F}_{10110000} &\approx 8 {\cal F}_{0} + {\Delta}{\cal F}_{ \overline{0} 1011 \overline{0} , \overline{0}1011 \overline{0} }  + {\Delta}{\cal F}_{ \overline{0} 1011 \overline{0} ,\overline{0} } \\
&=8 {\cal F}_{0} +  {\Delta}{\cal F}_{ \overline{0} 1011 \overline{0} ,\overline{0} }\\
&=50.510819938430132  \nonumber
\end{split}
\end{equation}
and the optimal result,
\begin{equation}\label{eq:Homoclinic decomposition action formula reshuffle numerical optimal}
\begin{split}
&{\cal F}_{11000010} \approx 8 {\cal F}_{0} + {\Delta}{\cal F}_{ \overline{0} 1100001 \overline{0} , \overline{0}11 \overline{0} } + {\Delta}{\cal F}_{ \overline{0} 1011 \overline{0} , \overline{0} 1 \overline{0} } \\
&=50.526729754916772\ .  \nonumber
\end{split}
\end{equation}
Comparing the results with Eq.~\eqref{eq:Homoclinic decomposition action formula reshuffle numerical periodic orbit action}, the relative error from the non-optimal partition is $3.1\times 10^{-4}$, and that of the optimal partition $5.9\times 10^{-6}$, a factor $190$ smaller.  

For the verification of the exact formula, substituting the optimal partition into Eq.~\eqref{eq:Homoclinic decomposition action formula reshuffle exact} leads to
\begin{equation}\label{eq:Homoclinic decomposition action formula reshuffle numerical exact}
\begin{split}
{\cal F}_{11000010}& = 50.526729754916772 - 0.000298551864113\\
& = 50.526431203052659\ . \nonumber
\end{split}
\end{equation}
Comparing with Eq.~\eqref{eq:Homoclinic decomposition action formula reshuffle numerical periodic orbit action}, the relative error of the exact formula is $7.9\times 10^{-11}$, as good as possible given the presence of propagation error.  

\section{Cycle expansion}
\label{Cycle expansion}

The cycle expansion \cite{Cvitanovic88,Cvitanovic89,ChaosBook} has been an important approach to determining various physical quantities, such as dynamical zeta functions and spectral determinants, in terms of the very few shortest periodic orbits.  In the expansion, a periodic orbit $\lbrace y \rbrace$ is grouped together with shorter pseudo-orbits whose full symbolic itineraries build up the itinerary of $\lbrace y \rbrace$, with the assumption that the action differences between $\lbrace y \rbrace$ and its decomposed pseudo-orbit decrease rapidly with increasing periods. Though it may be often true for highly chaotic systems, we show that a natural consequence of Eq.~\eqref{eq:Homoclinic decomposition action formula reshuffle approximate} yields a geometric result for the action differences, thus providing an analytic way to evaluate in which cases the action differences between orbits in the cycle expansions are small enough.            

\subsection{Action differences between periodic orbits and their decomposed pseudo-orbits}
\label{Action difference between periodic orbits and their decomposed pseudo-orbits}    
   
Consider an arbitrary periodic orbit $\lbrace y \rbrace \Rightarrow  \overline{\gamma} $, and its  partition into two substrings: $\gamma=\gamma_1\gamma_2$. In the cycle expansion, $\overline{\gamma}$ can be decomposed into a pseudo-orbit composed by $( \overline{\gamma_1} + \overline{\gamma_2} )$, with the action difference between them assumed vanishing in the original approaches. However, with the help of Eqs.~\eqref{eq:Homoclinic decomposition action formula reshuffle approximate} and \eqref{eq:Homoclinic decomposition action formula single approximate}, we can express the classical actions of $\overline{\gamma}$, $\overline{\gamma_1} $ and $\overline{\gamma_2}$ respectively:
\begin{equation}\label{eq:Cycle expansion long periodic orbit}
\begin{split}
&{\cal F}_{\gamma} = n_{\gamma}{\cal F}_{0} + {\Delta}{\cal F}_{ \overline{0}\gamma_1\gamma_2\overline{0} , \overline{0}\gamma_1\overline{0} } + {\Delta}{\cal F}_{ \overline{0}\gamma_2\gamma_1\overline{0} , \overline{0}\gamma_2\overline{0} }\\
& + O( e^{ -\mu (n_1 + L_2 + R_2) }  +  e^{ -\mu (n_2 + L_1 + R_1) } ) \ , 
\end{split}
\end{equation}
\begin{equation}\label{eq:Cycle expansion short periodic orbits 1}
{\cal F}_{\gamma_1} =n_1 {\cal F}_{0} + {\Delta}{\cal F}_{ \overline{0}\gamma_1\gamma_1\overline{0} , \overline{0}\gamma_1\overline{0} } + O(e^{-\mu (n_1 +L_1+R_1 )})
\end{equation}
and
\begin{equation}\label{eq:Cycle expansion short periodic orbits 2}
{\cal F}_{\gamma_2} =n_2 {\cal F}_{0} + {\Delta}{\cal F}_{ \overline{0}\gamma_2\gamma_2\overline{0} , \overline{0}\gamma_2\overline{0} }  + O(e^{-\mu (n_2 +L_2+R_2 )})\ .
\end{equation}
Subtracting Eqs.~\eqref{eq:Cycle expansion short periodic orbits 1} and \eqref{eq:Cycle expansion short periodic orbits 2} from Eq.~\eqref{eq:Cycle expansion long periodic orbit} gives an expression for the action difference between $\overline{\gamma}$ and its decomposed pseudo-orbit:
\begin{equation}\label{eq:Cycle expansion action difference formula previous}
\begin{split}
&{\cal F}_{ \gamma }  - ( {\cal F}_{\gamma_1} + {\cal F}_{\gamma_2} )\\
& ={\Delta}{\cal F}_{ \overline{0}\gamma_1\gamma_2\overline{0} , \overline{0}\gamma_1\overline{0} }  + {\Delta}{\cal F}_{ \overline{0}\gamma_2\gamma_1\overline{0} , \overline{0}\gamma_2\overline{0} }\\
& -{\Delta}{\cal F}_{ \overline{0}\gamma_1\gamma_1\overline{0} , \overline{0}\gamma_1\overline{0} } - {\Delta}{\cal F}_{ \overline{0}\gamma_2\gamma_2\overline{0} , \overline{0}\gamma_2\overline{0} }\\
& +O \Big( e^{ -\mu (n_1 + L_2 + R_2) }  +  e^{ -\mu (n_2 + L_1 + R_1) }  \\
& \qquad +e^{-\mu (n_1 +L_1+R_1 )} + e^{-\mu (n_2 +L_2+R_2 )} \Big) \ .
\end{split}
\end{equation}
Notice that, since
\begin{equation}
\begin{split}
&{\Delta}{\cal F}_{ \overline{0}\gamma_1\gamma_2\overline{0} , \overline{0}\gamma_1\overline{0} } - {\Delta}{\cal F}_{ \overline{0}\gamma_1\gamma_1\overline{0} , \overline{0}\gamma_1\overline{0} } ={\Delta}{\cal F}_{ \overline{0}\gamma_1\gamma_2\overline{0} , \overline{0}\gamma_1\gamma_1\overline{0} } \\
&{\Delta}{\cal F}_{ \overline{0}\gamma_2\gamma_1\overline{0} , \overline{0}\gamma_2\overline{0} } - {\Delta}{\cal F}_{ \overline{0}\gamma_2\gamma_2\overline{0} , \overline{0}\gamma_2\overline{0} } ={\Delta}{\cal F}_{ \overline{0}\gamma_2\gamma_1\overline{0} , \overline{0}\gamma_2\gamma_2\overline{0} }
\end{split}
\end{equation}
in fact, there is the simplified formula
\begin{equation}\label{eq:Cycle expansion action difference formula}
\begin{split}
&{\cal F}_{\gamma}  - ( {\cal F}_{\gamma_1} + {\cal F}_{\gamma_2} )\\
& ={\Delta}{\cal F}_{ \overline{0}\gamma_1\gamma_2\overline{0} , \overline{0}\gamma_1\gamma_1\overline{0} } + {\Delta}{\cal F}_{ \overline{0}\gamma_2\gamma_1\overline{0} , \overline{0}\gamma_2\gamma_2\overline{0} }\\
& +O \Big( e^{ -\mu (n_1 + L_2 + R_2) }  +  e^{ -\mu (n_2 + L_1 + R_1) }  \\
& \qquad +e^{-\mu (n_1 +L_1+R_1 )} + e^{-\mu (n_2 +L_2+R_2 )} \Big) 
\end{split}
\end{equation}
which expresses the action difference between $ \overline{\gamma} $ and $( \overline{\gamma_1}  + \overline{ \gamma_2 } )$ in terms of four homoclinic orbits: $ \overline{0}\gamma_1\gamma_2\overline{0}$, $ \overline{0}\gamma_1\gamma_1\overline{0}$, $ \overline{0}\gamma_2\gamma_1\overline{0}$ and $\overline{0}\gamma_2\gamma_2\overline{0}$, all constructed using the substrings of the periodic orbit. 

Furthermore, letting $h^{(\gamma_i\gamma_j)}_0 \Rightarrow \overline{0} \gamma_i . \gamma_j \overline{0}$ ($i,j=1,2$), a geometric alternative of Eq.~\eqref{eq:Cycle expansion action difference formula}, similar to Eq.~\eqref{eq:Homoclinic decomposition action formula reshuffle approximate geometric}, can be written as
\begin{equation}\label{eq:Cycle expansion action difference formula geometric}
\begin{split}
&{\cal F}_{\gamma}  - ( {\cal F}_{\gamma_1} + {\cal F}_{\gamma_2} )\\
& = {\cal A}^\circ_{SUSU[ h^{(\gamma_1\gamma_2)}_0 , h^{(\gamma_2\gamma_2)}_0, h^{(\gamma_2\gamma_1)}_0, h^{(\gamma_1\gamma_1)}_0 ]}\\
& +O \Big( e^{ -\mu (n_1 + L_2 + R_2) }  +  e^{ -\mu (n_2 + L_1 + R_1) }  \\
& \qquad +e^{-\mu (n_1 +L_1+R_1 )} + e^{-\mu (n_2 +L_2+R_2 )} \Big) 
\end{split}
\end{equation}
which calculates the action difference as a curvy parallelogram bounded by stable and unstable manifolds.  The quality of the cycle expansion depends on the size of the areas ${\cal A}^\circ_{SUSU[....]}$, which are typically small if a long orbit is split in the middle. However, it is conceivable that one might include correction terms, such as the area term in Eq.~\eqref{eq:Cycle expansion action difference formula geometric}, into the approximation to achieve a better result, although a more elaborate resummation scheme is needed for that purpose.  

For general cases where $\gamma$ is exceedingly long, further partitioning is possible by repeated use of Eq.~\eqref{eq:Cycle expansion action difference formula}, and therefore the action difference between a long periodic orbit and its decomposed pseudo-orbit composed by multiple substrings can be obtained.  This adds further area correction terms.  At a minimum, the magnitudes of the error terms imply a practical limit to further decreasing lengths $n_i$ of the substrings $\gamma_i$, which may depnd on the error tolerance of specific problems (or the value of $\hbar$). 
 
\subsection*{Optimal decomposition and numerical verification}
\label{Optimal decomposition and numerical verification}

The results just presented allow for arbitrary decompositions of any given periodic orbit $ \overline{\gamma} $: $\gamma=\gamma_1\gamma_2$, and express the action difference between $ \overline{\gamma} $ and $( \overline{\gamma_1}  +  \overline{\gamma_2} )$ in terms of specific homoclinic relative actions or phase space areas.  However, it is worth comparing optimal and non-optimal partitions for the cycle orbits.  The optimal partitions $\gamma^{\prime}=\gamma^{\prime}_1\gamma^{\prime}_2$ maximize:
\begin{equation}\label{eq:Cycle expansion optimal decomposition maximize}
\begin{split} 
\min\big(n_1 &+ L_2 + R_2\ ,\ n_2 + L_1 + R_1\ ,\\
 &n_1 +L_1+R_1\ ,\ n_2 +L_2+R_2 \big) \ .
 \end{split}
\end{equation} 
Again, the optimal decomposition may be not unique for a periodic orbit. 
  
As a numerical verification, consider the period-$8$ orbit $\lbrace y \rbrace$ ($\gamma=10110000$) from Sec.~\ref{Optimal partition and numerical verification}.  A non-optimal decomposition is given again by $\gamma_1=1011$ and $\gamma_2=0000$.  The $\overline{1011} $  orbit has $y_0 = (3.1622776601683795,1.9171449292276370)$, whereas the $ \overline{0000}$ orbit is just four times the fixed point orbit. Substituting their symbolic strings into Eq.~\eqref{eq:Cycle expansion action difference formula} yields
\begin{equation}\label{eq:Cycle expansion non optimal example theoretical}
\begin{split}
&{\cal F}_{10110000}  - ({\cal F}_{1011} + {\cal F}_{0000} )\\
&\approx {\Delta}{\cal F}_{ \overline{0}1011\overline{0} , \overline{0}10111011\overline{0} } + {\Delta}{\cal F}_{ \overline{0}1011\overline{0} , \overline{0} }\ .
\end{split}
\end{equation}
The difference is given by
\begin{equation}\label{eq:Cycle expansion non optimal example periodic action difference}
\begin{split}
{\cal F}_{10110000} & - ({\cal F}_{1011} + {\cal F}_{0000} )\\
&=-47.264193841143715 \ ,
\end{split}
\end{equation}
which would actually be a nonsensical split for the cycle expansion as one wants this difference already to be small and it is on the scale of the action for the full orbit.  The correction area is
\begin{equation}\label{eq:Cycle expansion non optimal example homoclinic relative action}
\begin{split}
{\Delta}{\cal F}_{ \overline{0}1011\overline{0} , \overline{0}10111011\overline{0} } & + {\Delta}{\cal F}_{ \overline{0}1011\overline{0} , \overline{0} }\\
& =-47.318648376144232\ ,
\end{split}
\end{equation}
leaving an error relative to the full periodic orbit action as $1.1 \times 10^{-3}$.   

Next, consider the optimal decomposition $\gamma^{\prime}_1=0010$ and $\gamma^{\prime}_2=1100$. The $\overline{0010}$ orbit  $y_0=(-4.0403657409121712,-3.1622776601683793)$, and the $\overline{1100}$ orbit has $y_0=(3.1622776601683793,-3.1622776601683793)$.  Substituting them into Eq.~\eqref{eq:Cycle expansion action difference formula} yields
\begin{equation}\label{eq:Cycle expansion optimal example theoretical}
\begin{split}
&{\cal F}_{ 00101100 }-({\cal F}_{0010}+{\cal F}_{1100})\\
&\approx  {\Delta}{\cal F}_{ \overline{0}1011\overline{0} , \overline{0}10001\overline{0} } + {\Delta}{\cal F}_{ \overline{0}1100001\overline{0} , \overline{0}110011\overline{0} }\ \  .
\end{split}
\end{equation}
This gives
\begin{equation}\label{eq:Cycle expansion optimal example periodic action difference}
{\cal F}_{ 00101100 }-({\cal F}_{0010}+{\cal F}_{1100})= 8.3627290714575508 \nonumber
\end{equation}
which is far superior to the previous case as the area is much smaller.  It is still significant though.  The area correction is
\begin{equation}\label{eq:Cycle expansion optimal example homoclinic relative action}
\begin{split}
{\Delta}{\cal F}_{ \overline{0}1011\overline{0} , \overline{0}10001\overline{0} } +& {\Delta}{\cal F}_{ \overline{0}1100001\overline{0} , \overline{0}110011\overline{0} }\\
& = 8.3635870750844319\ ,
\end{split}
\end{equation}
leaving a relative error with respect to the full action of $1.7 \times 10^{-5}$, a significant improvement compared with the non-optimal decomposition, as expected.

\section{Conclusions}
\label{Conclusions}
The relations between classical orbit sets play important roles in both classical and quantum chaotic dynamics.  The relations given here can be used as a starting point for understanding the connections betweeen homoclinic and periodic orbits, action correlations, corrections to cycle expansions, and symmetries, such as the role of Richter-Sieber pairs in time reversal invariant systems.  Controlled estimates of errors were given for various approximations.  Also, note that individual periodic orbit actions can be calculated from generating functions, but the numerical computations suffer from sensitive dependence on initial conditions therefore being prohibitive for long orbits. By relating them to homoclinic orbits, which can be stably calculated relying on the structural stability of stable and unstable manifolds, they become stably calculable as well.

Organizing the orbits with symbolic dynamics, we have determined periodic orbit actions using homoclinic orbits constructed from either the entire symbolic string or partitioned substrings, and derived both exact and approximate action formulae.  Although the exact formulae require the numerical determination of the periodic orbit points, the approximate formulae only require the calculation of homoclinic orbits, which is much simpler~\cite{Li17}. The errors associated with the approximate formulae scale down exponentially fast with increasing periods, making them almost exact for long orbits. Explicit action relations amongst periodic orbits also come as natural consequences, which turns the often empirical or statistical account of action correlations between periodic orbits into an analytic study of either homoclinic relative actions or phase space areas bounded by invariant manifolds, linking classical entities such as homoclinic tangles to the quantal spectral quantities of chaotic systems.        

The analytic scheme developed here provides universal expressions for the action relations between either homoclinic and periodic orbits, periodic orbit pairs, or periodic orbits and their decomposed pseudo-orbits on the microscopic level. It is conceivable that this microscopic formula, when paired with a macroscopic counting scheme~\cite{Gutkin13} should enable efficient semiclassical resummations on both the analytic~\cite{Muller05} and numerical aspects.

\appendix

\section{HORSESHOE, MARKOV PARTITIONS AND SYMBOLIC DYNAMICS}

Symbolic dynamics provides a powerful technique, i.e.~the topological description of orbits in chaotic systems~\cite{Hadamard1898,Birkhoff27a,Birkhoff35,Morse38}. Perhaps the most famous model that demonstrates its elegance is the horseshoe map~\cite{Smale63,Smale80}, a two-dimensional diffeomorphism possessing an invariant Cantor set, which is topologically conjugate to a Bernoulli shift on symbolic strings composed by ``$0$"s and ``$1$"s. A numerical realization of the horseshoe is the area-preserving H\'{e}non map \cite{Henon76} defined on the phase plane $(q,p)$, which is the simplest polynomial automorphism giving rise to chaotic dynamics~\cite{Friedland89}:
\begin{equation}\label{eq:Henon map}
\begin{split}
&p_{n+1}=q_n\\
&q_{n+1}=a-q_{n}^2-p_n.
\end{split}
\end{equation}
It follows from the work in Ref.~\cite{Devaney79} that for sufficiently large parameter values of $a$ the H\'{e}non map is topologically conjugate to a horseshoe map, therefore possessing a hyperbolic invariant set of orbits labeled by binary symbolic codes; see Chapters 23 and 24 of Ref.~\cite{Wiggins03} for a brief review of the Smale horseshoe and the corresponding symbolic dynamics. 
\label{Markov partition}

To visualize the action of the mapping $M$ (e.g. Eq.\eqref{eq:Henon map}) on the homoclinic tangle, let us consider the closed region $\cal{R}$ in Fig.~\ref{fig:horseshoe}, bounded by loop $\mathcal{L}_{USUS[x,g_{-1},h_0,g_0]}$, where $\mathcal{L}_{USUS[x,g_{-1},h_0,g_0]}=U[x,g_{-1}]+S[g_{-1},h_0]+U[h_0,g_0]+S[g_0,x]$. Under the mapping $M$, the trapezoid-shaped $\cal{R}$ is compressed along the stable direction and stretched along the unstable direction, and folded back to partially overlap with itself, with the vertical strips $V_0$ and $V_1$ mapped into the horizontal strips $H_0$ and $H_1$, respectively. Similarly, the inverse mapping $M^{-1}$ stretches $\cal{R}$ along the stable direction and fold back, with the horizontal strips $H_0$ and $H_1$ mapped into $V_0$ and $V_1$, respectively. Therefore, points in region $E_0$ bounded by $\mathcal{L}_{USUS[g_{-2},h_{-1},h^{\prime}_{-1},g^{\prime}_{-1}]}$ are mapped outside $\cal{R}$ into $E_1$ bounded by $\mathcal{L}_{USUS[g_{-1},h_{0},h^{\prime}_{0},g^{\prime}_{0}]}$ under one iteration. For open systems such as the H\'{e}non map, any point outside $\cal{R}$  never returns and escapes to infinity; there is a similar construction for inverse time.  Of great structural significance is the non-wandering set $\Omega$ of phase-space points $z$ that stay inside $\cal{R}$ for all iterations~\cite{ChaosBook,Wiggins03}:        
\begin{equation}\label{eq:Nonwandering set}
\Omega=\big\lbrace z:z\in \bigcap_{n=-\infty}^{\infty}M^{n}(\cal{R}) \big\rbrace.
\end{equation}  
In particular, we focus on the homoclinic and periodic points that belong to $\Omega$.  

Using the closed regions $V_0$ and $V_1$ in Fig.~\ref{fig:horseshoe} as Markov generating partition for the symbolic dynamics, every point $z_0$ in $\Omega$ can be labeled by an infinite symbolic string of $0$'s and $1$'s:
\begin{equation}\label{eq:symbolic code}
z_0 \Rightarrow \cdots s_{-2}s_{-1}.s_{0}s_{1}s_{2}\cdots
\end{equation}
where each digit $s_n$ in the symbol denotes the region that $M^{n}(z_0)$ lies in: $M^{n}(z_0) = z_n \in V_{s_{n}}$, $s_n \in \lbrace 0,1\rbrace$. In that sense, the symbolic code gives an ``itinerary" of $z_0$ under successive forward and backward iterations, in terms of the regions $V_0$ and $V_1$ that each iteration lies in. The semi-infinite segment ``$s_{0}s_{1}s_{2}\cdots$" (resp. ``$\cdots s_{-2}s_{-1}$") from the symbolic code is referred to as the $\mathit{head}$ (resp. $\mathit{tail}$) of the orbit with initial condition $z_0$~\cite{Hagiwara04}, and the decimal point separating the head and the tail denotes the region ($V_{s_0}$) that the current iteration $z_0$ belongs to. Let $\Sigma$ denote the symbolic space of all such bi-infinite symbolic strings. Strings in $\Sigma$ are then in 1-to-1 correspondence with points in $\Omega$, and the mapping $M$ in phase space is topological conjugate to a Bernoulli shift in the symbolic space.  Therefore, forward iterations of $z_0$ move its decimal point towards the right side of the symbolic string, and backward iterations move it towards the left side. 

\begin{figure}
 \subfigure{
   \label{fig:Markov_1}
   \includegraphics[width=5.5cm]{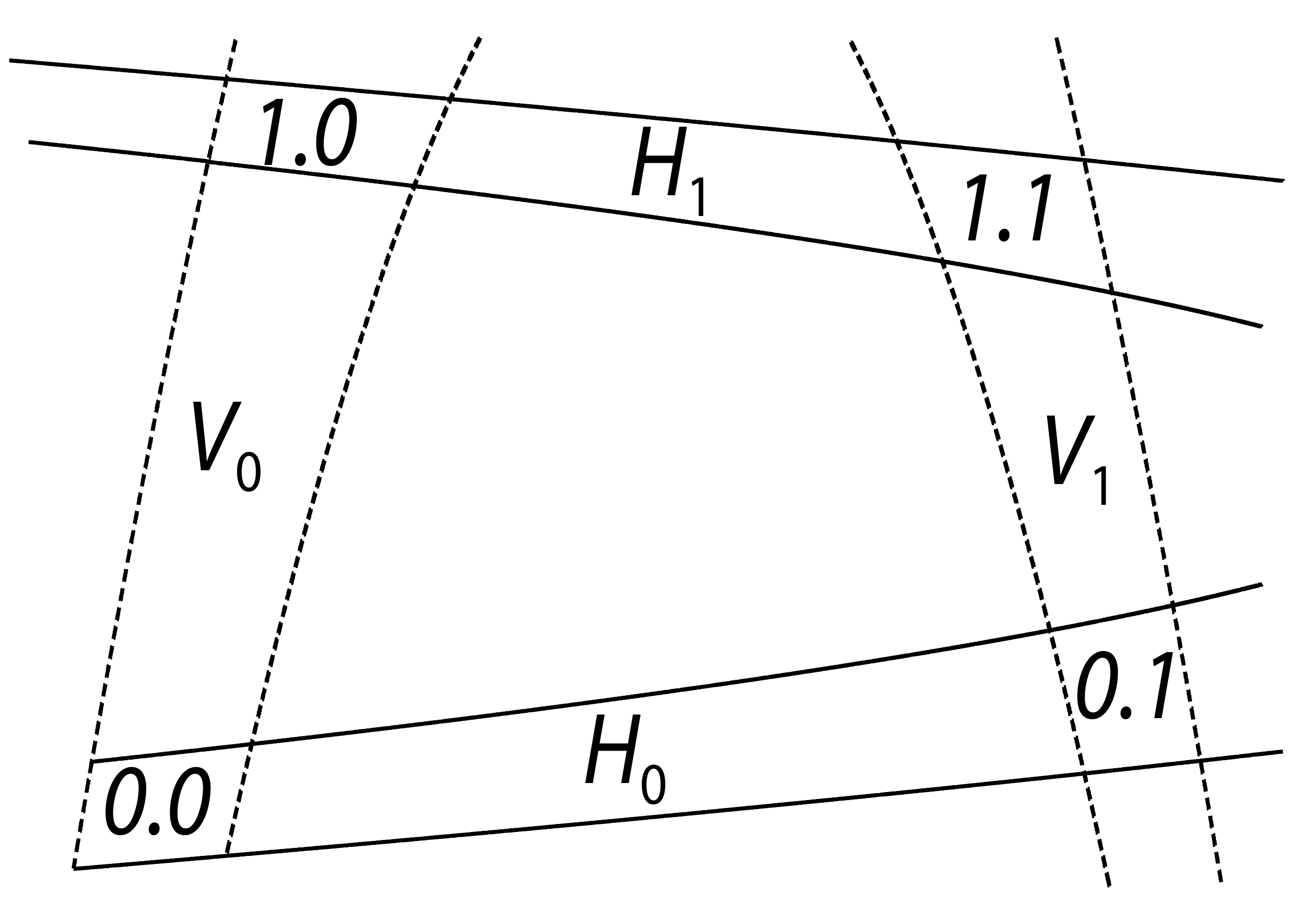}}
 \subfigure{
   \label{fig:Markov_2}
   \includegraphics[width=5.5cm]{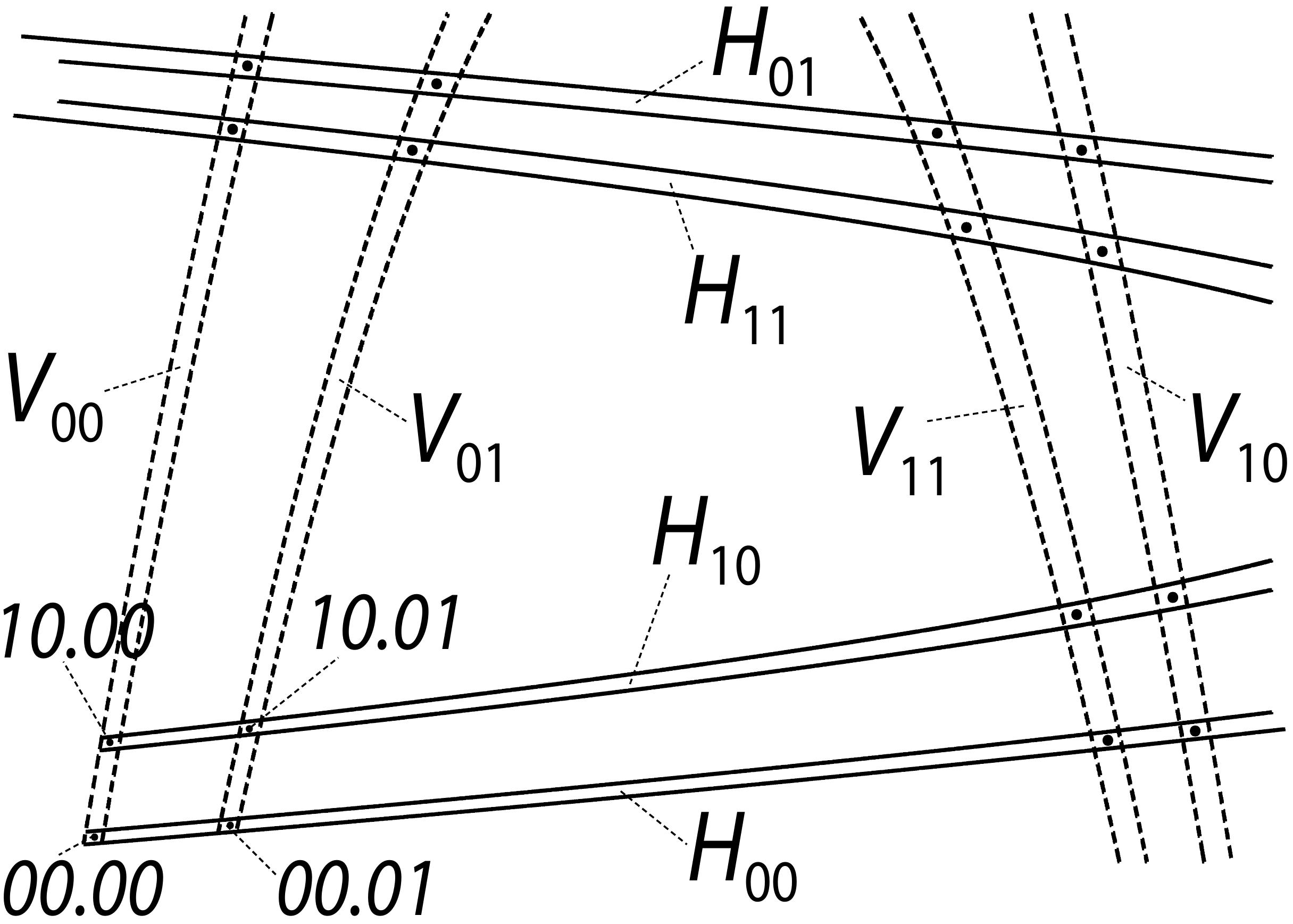}}
\caption{Markov partitions constructed in the H\'{e}non map. Upper panel: The $V_{s_{0}}$ and $H_{s_{-1}}$ regions corresponds to the same regions in Fig.~\ref{fig:horseshoe}. The four cells $H_{s_{-1}}\cap V_{s_0} \Rightarrow s_{-1}.s_{0}$ are the Markov partitions of lengths $2$. Lower panel: Markov partitions of length $4$. The horizontal and vertical strips are created as $H_{s_{-2}s_{-1}}=M(H_{s_{-2}})\cap H_{s_{-1}}$ and $V_{s_0 s_1}=V_{s_0} \cap M^{-1}(V_{s_1})$. The $H$ and $V$ strips intersect at sixteen cells $H_{s_{-2}s_{-1}} \cap V_{s_0 s_1} \Rightarrow s_{-2} s_{-1}.s_0 s_1$, as indicated by a black dot inside each of them. For the sake of clarity, we only explicitly labeled four cells in the lower left corner. Any point from $\Omega$ with symbolic string of fixed central block $\cdots s_{-2} s_{-1}.s_0 s_1 \cdots$ must either locate inside or on the boundary of the $s_{-2}s_{-1}.s_0 s_1$ cell. The sizes of the cells shrink exponentially with increasing string lengths.  } 
\label{fig:Markov}
\end{figure}         

Besides elegant topological conjugacy, the symbolic strings also contain information about the location of points in phase space. Following a standard procedure \cite{Wiggins88}, Subsequent Markov partitions \cite{Bowen75,Gaspard98} can be constructed from the generating partitions $[V_0,V_1]$, which specifies the phase-space regions that points with certain central blocks of fixed lengths must locate within. Starting from $V_0$ and $V_1$, define recursively an ever-shrinking family of vertical strips $V_{s_0\cdots s_{n-1}}$ in phase space, such that:
\begin{equation}\label{eq:Markov partition vertical strips}
V_{s_0\cdots s_{n-1}}\equiv V_{s_0} \bigcap M^{-1}(V_{s_1\cdots s_{n-1}})
\end{equation}  
where $s_{i}\in \lbrace 0,1\rbrace$ for $i=0,\cdots,n-1$. Similarly, starting from $H_0$ and $H_1$, an ever-shrink family of horizontal strips $H_{s_{-n}\cdots s_{-1}}$ can be defined:
\begin{equation}\label{eq:Markov partition horizontal strips}
H_{s_{-n}\cdots s_{-1}} \equiv M(H_{s_{-n}\cdots s_{-2}}) \bigcap H_{s_{-1}} 
\end{equation}  
where $s_{-j}\in \lbrace 0,1\rbrace$ for $j=1,\cdots,n$. The horizontal and vertical strips intersect at curvy ``rectangular" cells, which can be labeled by a finite string of symbols: 
\begin{equation}\label{eq:Markov partition cells}
H_{s_{-n}\cdots s_{-1}} \bigcap V_{s_0\cdots s_{n-1}} \Rightarrow s_{-n}\cdots s_{-1}.s_{0}\cdots s_{n-1} 
\end{equation}

These cells are Markov partitions of central block lengths $2n$, in the sense that any point from $\Omega$ with coinciding central blocks $s_{-n}\cdots s_{-1}.s_{0}\cdots s_{n-1}$ must locate inside (or on the boundary of) the corresponding cell. Shown in the upper and lower panels of Fig.~\ref{fig:Markov} are two examples of Markov partitions of lengths $2$ and $4$, respectively, numerically generated from the H\'{e}non map. Take the cell $10.01$ from the lower panel as example, any point with symbolic string of the form: $\cdots s_{-4}s_{-3} 10.01 s_2 s_3 \cdots$ must either locate inside or on the boundary of $10.01$. 

\begin{figure}[ht]
\centering
{\includegraphics[width=6.5cm]{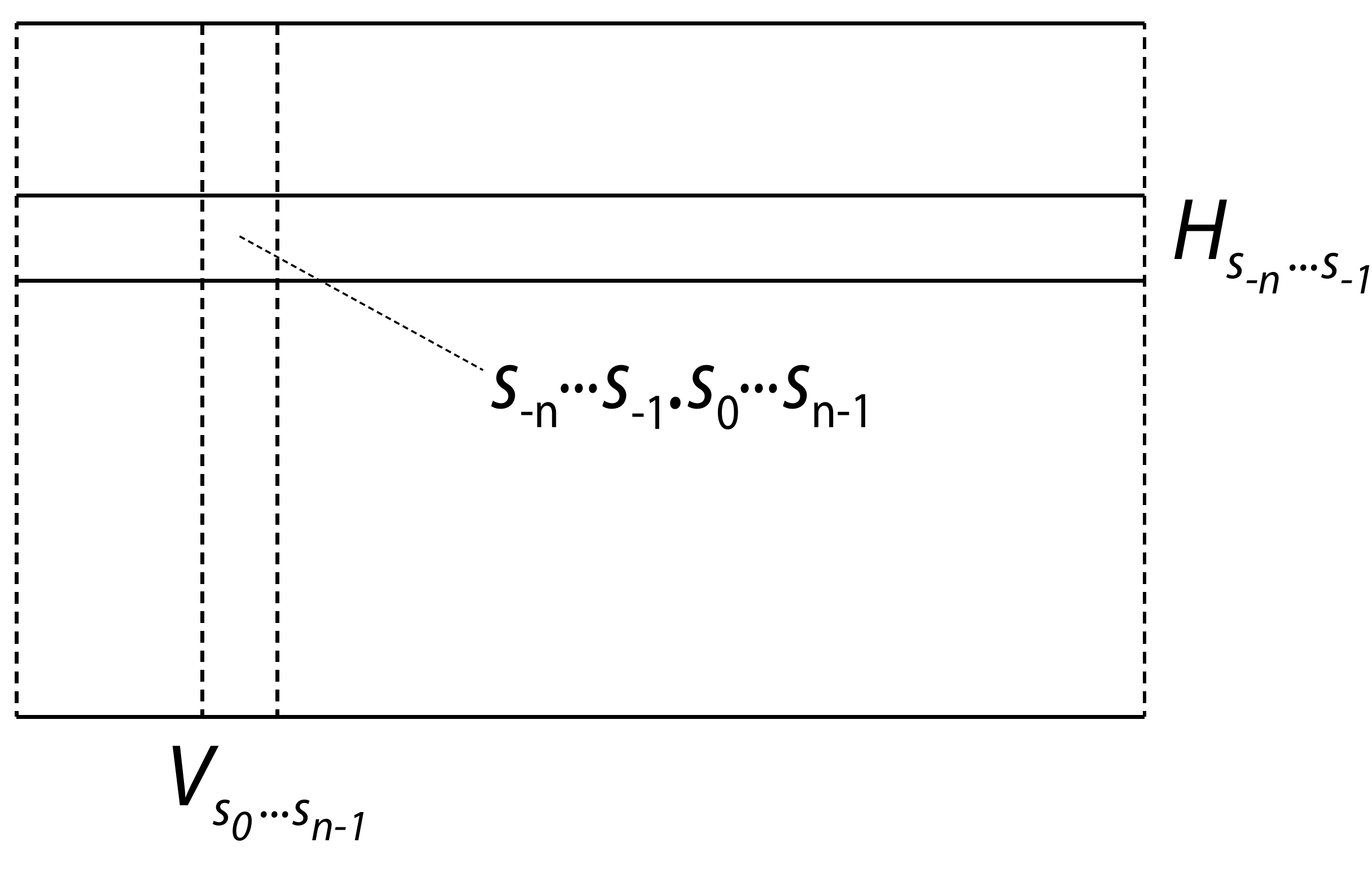}}
 \caption{(Schematic) The widths of $H_{s_{-n}\cdots s_{-1}}$ and $V_{s_0\cdots s_{n-1}}$ are $\sim O(e^{-n\mu})$, so the cell area of $s_{-n} \cdots s_{-1}.s_0 \cdots s_{n-1}$ is $\sim O(e^{-2n\mu})$. }
\label{fig:Shrinking_Cells}
\end{figure}  

Closeness between two symbolic strings imply closeness between the corresponding points in phase space. Due the compressing and stretching nature of the horseshoe map, the widths of the horizontal and vertical strips becomes exponentially small with increasing block lengths, and so do the cell areas they intersect. Without loss of generality, we assume, in Fig.~\ref{fig:horseshoe}, that the area ${\cal A}^{\circ}_{SUSU[x,g_0,h_0,g_{-1}]}$ is of order $\sim O(1)$. Then the resulting area of the cell $s_{-n}\cdots s_{-1}.s_{0}\cdots s_{n-1}$ is of order $\sim O(e^{-2n\mu})$, where $\mu$ is the Lyapunov exponent of the system, an exponentially small area for large $n$ values. This geometry \cite{Wiggins88} is shown by Fig.~\ref{fig:Shrinking_Cells}. Therefore, any two points from $\Omega$ with identical central blocks of length $2n$ must locate in the same exponentially small cell. Consider twp points $h \Rightarrow \cdots s_{-n} \cdots s_{-1}.s_0 \cdots s_n \cdots$ and $h^{\prime} \Rightarrow \cdots s^{\prime}_{-n} \cdots s^{\prime}_{-1}.s^{\prime}_0 \cdots s^{\prime}_n \cdots$, if $h$ and $h^{\prime}$ agree on a central block of length $2n$, i.e., $s^{\prime}_{-n} \cdots s^{\prime}_{-1}.s^{\prime}_0 \cdots s^{\prime}_{n-1}=s_{-n}\cdots s_{-1}.s_0\cdots s_{n-1}$, they must both located in same cell labeled by $s_{-n} \cdots s_{-1}.s_0 \cdots s_{n-1}$
\begin{equation}\label{eq:Matching central block lengths area estimate}
h,h^{\prime} \in H_{s_{-n}\cdots s_{-1}} \bigcap V_{s_0\cdots s_{n-1}} \Rightarrow s_{-n}\cdots s_{-1}.s_0\cdots s_{n-1}
\end{equation}
the area of which is $\sim O(e^{-2n\mu})$. Therefore, by specifying longer and longer central block lengths of a point's symbolic string, we can narrow down its possible location in phase space with smaller and smaller cells from the Markov partition.

\section{MACKAY-MEISS-PERCIVAL ACTION PRINCIPLE}
\label{MacKay-Meiss-Percival}

\begin{figure}[ht]
\centering
{\includegraphics[width=8cm]{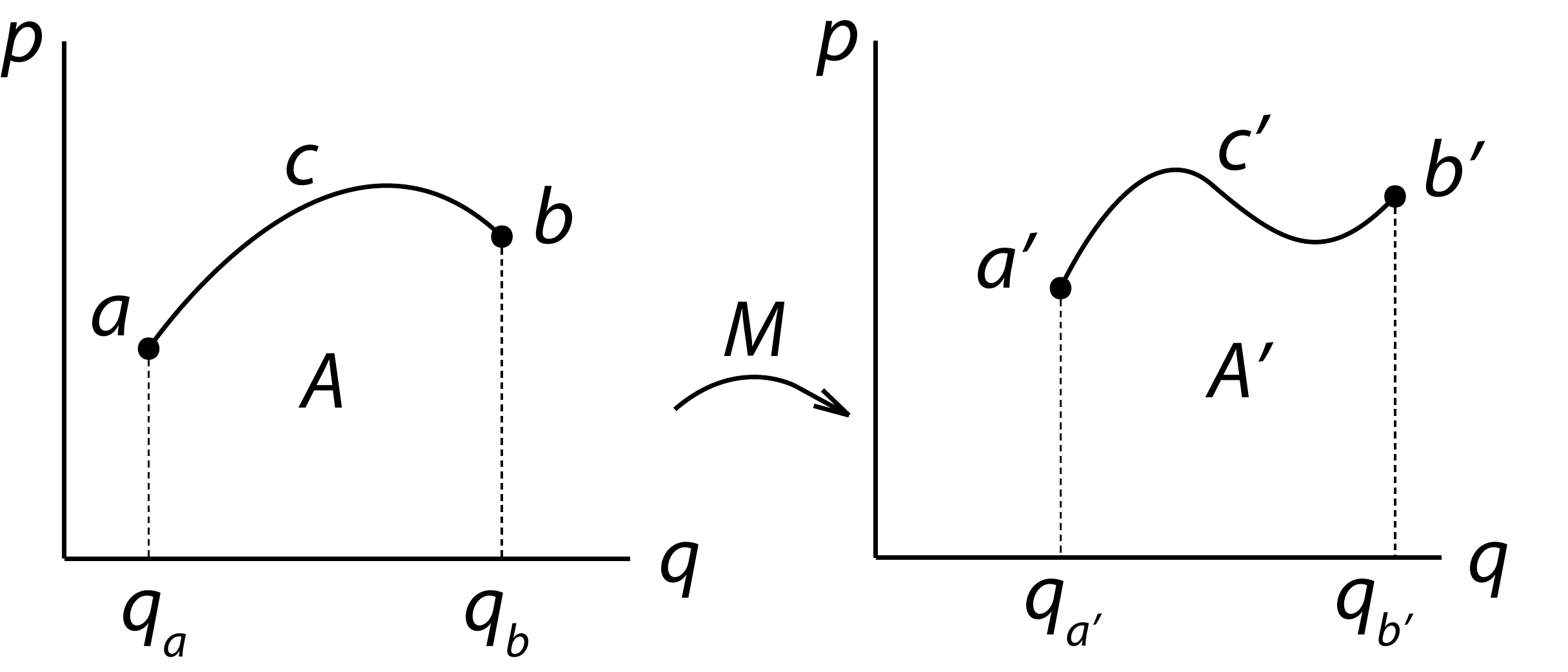}}
 \caption{$a$ and $b$ are arbitrary points and $c$ is a curve connecting them.  $a'=M(a)$, $b'=M(b)$ and $c'=M(c)$.  Then: $A'-A=F(q_{b},q_{b'})-F(q_{a},q_{a'})$. }
\label{fig:Area_under_curve}
\end{figure}  

The MacKay-Meiss-Percival action principle discussed in this section was first developed in \cite{MacKay84a} for transport theory.  A comprehensive review can be found in \cite{Meiss92}. Generalization of the original principle beyond the ``twist" and area-preserving conditions is discussed in \cite{Easton91}, and we only give a brief outline of the theory in this appendix. Shown in Fig.~\ref{fig:Area_under_curve} are two arbitrary points $a=(q_{a},p_{a})$, $b=(q_{b},p_{b})$ and their images $a'=M(a)$, $b'=M(b)$.  Let $c$ be an arbitrary curve connecting $a$ and $b$, which is mapped to a curve $c'=M(c)$ connecting $a'$ and $b'$.  Let $A$ and $A'$ denote the algebraic area under $c$ and $c'$ respectively.  Then the difference between these areas is
\begin{equation}\label{eq:Meiss92}
\begin{split}
A'-A&=\int_{c'}p\mathrm{d}q-\int_{c}p\mathrm{d}q\\
&=F(q_{b},q_{b'})-F(q_{a},q_{a'})
\end{split}
\end{equation}
i.e., the difference between the two algebraic areas gives the difference between the action functions for one iteration of the map. 

\begin{figure}[ht]
\centering
{\includegraphics[width=4cm]{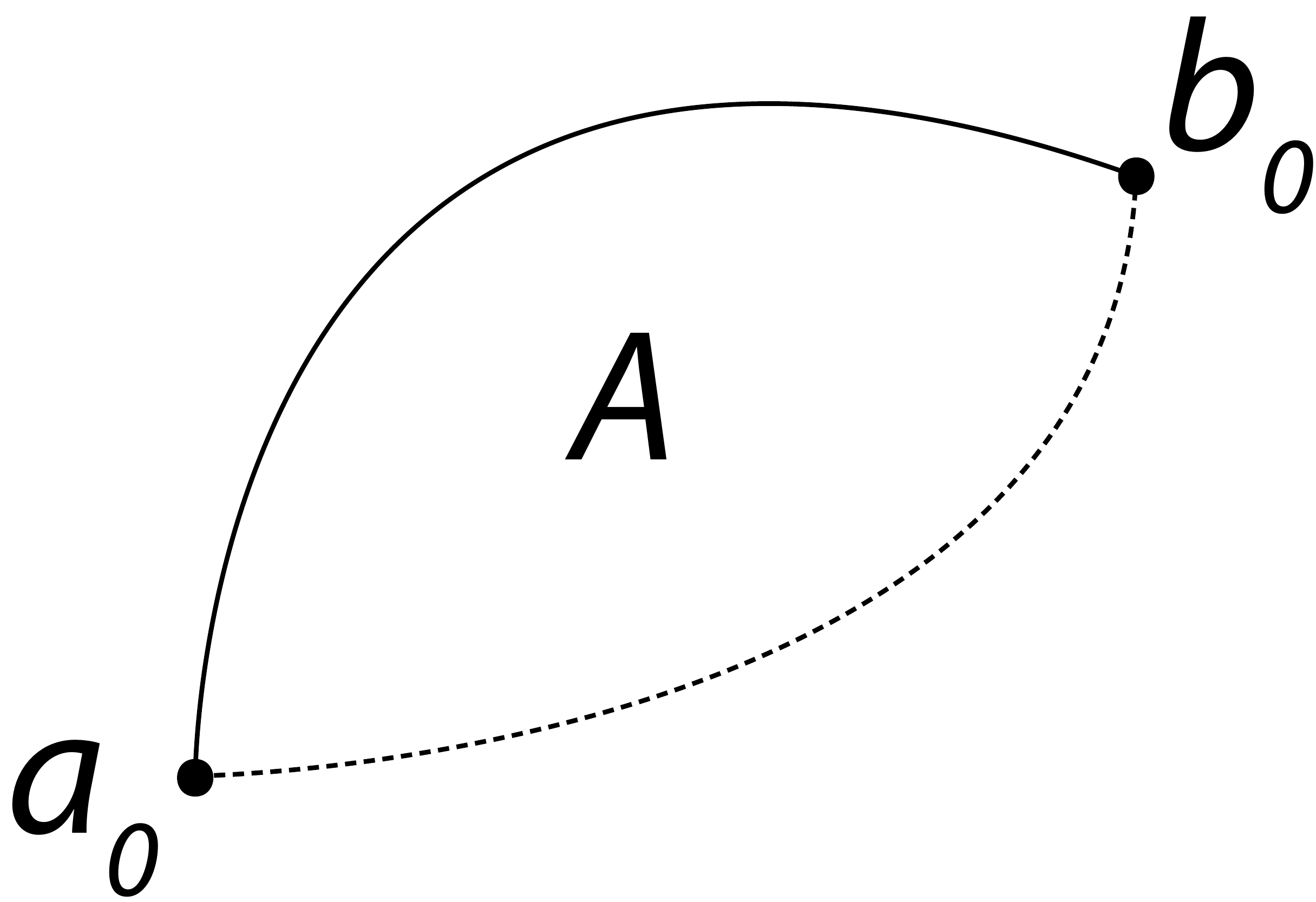}}
 \caption{$a_{0}$ and $b_{0}$ is a homoclinic pair.  They are connected by an unstable segment $U[a_{0},b_{0}]$ (solid) and a stable segment $S[b_{0},a_{0}]$ (dashed).  Then the action difference between the homoclinic orbit pair is $\Delta {\cal F}_{\lbrace b_0\rbrace \lbrace a_0\rbrace}=A$.    }
\label{fig:Homoclinic_pair_action}
\end{figure}

Starting from this, MacKay $\mathit{et}$ $\mathit{al.}$ \cite{MacKay84a} derived a formula relating the action difference between a pair of homoclinic orbits to the phase space area of a region bounded by stable and unstable manifolds, as demonstrated by Fig.~\ref{fig:Homoclinic_pair_action}.
In this Figure, $a_{0}$ and $b_{0}$ is a pair of homoclinic points:
\begin{equation}\label{eq:asymptotic pair}
a_{\pm\infty} \to b_{\pm\infty}\ .
\end{equation}
There exist unstable and stable manifolds connecting the two points shown by the solid and dashed curves.  Those manifolds could be the manifolds of other fixed points, or manifolds associated with $a_{0}$ and $b_{0}$ themselves.  Let $U[a_{0},b_{0}]$ and $S[b_{0},a_{0}]$ be the corresponding segments, we first apply Eq.~\eqref{eq:Meiss92} repeatedly to the semi-infinite pair of homoclinic orbit segments $\lbrace a_{-\infty},\cdots,a_{0}\rbrace$ and $\lbrace b_{-\infty},\cdots,b_0\rbrace$, and get:
\begin{equation}\label{eq:Homoclinic action difference infinite past appendix}
\begin{split}
&\sum_{n=-\infty}^{0} [F(b_{n-1},b_n)-F(a_{n-1},a_n)]\\
&=\int\limits_{U[a_{0},b_{0}]}p\mathrm{d}q-\int\limits_{U[a_{-\infty},b_{-\infty}]}p\mathrm{d}q=\int\limits_{U[a_{0},b_{0}]}p\mathrm{d}q
\end{split}
\end{equation}
where $\int\limits_{U[a_{-\infty},b_{-\infty}]}p\mathrm{d}q=0$ since $a_{-\infty} \to b_{-\infty}$. Similarly for the semi-infinite pairs $\lbrace a_{0},\cdots,a_{\infty}\rbrace$ and $\lbrace b_{0},\cdots,b_{\infty}\rbrace$ we have:
\begin{equation}\label{eq:Homoclinic action difference infinite future appendix}
\begin{split}
&\sum_{n=0}^{\infty} [F(b_{n},b_{n+1})-F(a_{n},a_{n+1})]\\
&=\int\limits_{S[a_{\infty},b_{\infty}]}p\mathrm{d}q-\int\limits_{S[a_{0},b_{0}]}p\mathrm{d}q=\int\limits_{S[b_{0},a_{0}]}p\mathrm{d}q\ .
\end{split}
\end{equation}
Adding up Eqs.~\eqref{eq:Homoclinic action difference infinite past appendix} and \eqref{eq:Homoclinic action difference infinite future appendix} we have:
\begin{equation}\label{eq:Homoclinic action difference appendix}
\begin{split}
\Delta {\cal F}_{\lbrace b_0\rbrace \lbrace a_0\rbrace}&=\sum_{n=-\infty}^{\infty}[F(b_n,b_{n+1}) - F(a_n, a_{n+1})]\\
&=\int\limits_{U[a_{0},b_{0}]}p\mathrm{d}q+\int\limits_{S[b_{0},a_{0}]}p\mathrm{d}q=A
\end{split}
\end{equation}
where $A$ denotes the area shown in Fig.~\ref{fig:Homoclinic_pair_action}.

\section{DERIVATION OF EQ.~\eqref{eq:Homoclinic decomposition action formula reshuffle exact}}
\label{Derivation of Equation}

This appendix contains a detailed derivation of Eq.~\eqref{eq:Homoclinic decomposition action formula reshuffle exact}, which follows a similar, but more elaborate process compared with those in Sec.~\ref{Action formula with entire periodic orbit strings}.  As in Sec.~\ref{Action formula with reshuffled substrings}, let the periodic orbit be $\lbrace y \rbrace \Rightarrow  \overline{\gamma} $, and its symbolic code partitioned into: $\gamma = \gamma_1 \gamma_2= \gamma^{-}_1 \gamma^{+}_1 \gamma^{-}_2 \gamma^{+}_2$, with lengths $n_{\gamma}=n_1+n_2= n^{-}_1 + n^{+}_1 + n^{-}_2 + n^{+}_2$. Its classical action ${\cal F}_{\gamma}$ can be extracted with the help of four auxiliary homoclinic orbits
\begin{equation}\label{eq:Accompanying homoclinic orbits reshuffled substrings appendix}
\begin{split}
   & \lbrace h^{(\gamma_1)}_0 \rbrace \Rightarrow  \overline{0} \gamma_1 \overline{0}  =  \overline{0} \gamma^{-}_1 \gamma^{+}_1 \overline{0} \\
   & \lbrace h^{(\gamma_2)}_0 \rbrace \Rightarrow \overline{0} \gamma_2 \overline{0} = \overline{0} \gamma^{-}_2 \gamma^{+}_2 \overline{0} \\
   & \lbrace h^{(\gamma_1 \gamma_2)}_0 \rbrace \Rightarrow  \overline{0} \gamma_1 \gamma_2 \overline{0} =  \overline{0} \gamma^{-}_1 \gamma^{+}_1 \gamma^{-}_2 \gamma^{+}_2 \overline{0} \\
   & \lbrace h^{(\gamma_2 \gamma_1)}_0 \rbrace \Rightarrow \overline{0} \gamma_2 \gamma_1 \overline{0}  =  \overline{0} \gamma^{-}_2 \gamma^{+}_2 \gamma^{-}_1 \gamma^{+}_1 \overline{0} \ .
  \end{split}
\end{equation}
Furthermore, set the zero subscript homoclinic points along their respective orbits
\begin{equation}\label{eq:Acompanying homoclinic orbit points setting appendix}
\begin{split}
& h^{(\gamma_1)}_0 \Rightarrow \overline{0} \gamma^{-}_1 . \gamma^{+}_1 \overline{0} \\
& h^{(\gamma_2)}_0 \Rightarrow \overline{0} \gamma^{-}_2 . \gamma^{+}_2 \overline{0} \\
& h^{(\gamma_1 \gamma_2)}_0 \Rightarrow \overline{0} \gamma^{-}_1 \gamma^{+}_1 . \gamma^{-}_2 \gamma^{+}_2 \overline{0}\\
& h^{(\gamma_2 \gamma_1)}_0 \Rightarrow \overline{0} \gamma^{-}_2 \gamma^{+}_2 . \gamma^{-}_1 \gamma^{+}_1 \overline{0}
\end{split}
\end{equation} 
from which it follows that
\begin{equation}\label{eq:Acompanying homoclinic orbit points setting 2 appendix}
\begin{split}
& h^{(\gamma_1 \gamma_2)}_{ -n^{+}_1 } \Rightarrow \overline{0} \gamma^{-}_1 . \gamma^{+}_1 \gamma^{-}_2 \gamma^{+}_2 \overline{0} \\
& h^{(\gamma_1 \gamma_2)}_{ n^{-}_2 } \Rightarrow \overline{0} \gamma^{-}_1 \gamma^{+}_1 \gamma^{-}_2 . \gamma^{+}_2 \overline{0} \\
& h^{(\gamma_2 \gamma_1)}_{ -n^{+}_2 } \Rightarrow \overline{0} \gamma^{-}_2 . \gamma^{+}_2 \gamma^{-}_1 \gamma^{+}_1 \overline{0} \\
& h^{(\gamma_2 \gamma_1)}_{ n^{-}_1 } \Rightarrow \overline{0} \gamma^{-}_2 \gamma^{+}_2 \gamma^{-}_1 . \gamma^{+}_1 \overline{0}
\end{split}
\end{equation} 
where the points in Eq.~\eqref{eq:Acompanying homoclinic orbit points setting 2 appendix} are just images of those in Eq.~\eqref{eq:Acompanying homoclinic orbit points setting appendix} under respective iterations. Of interest here are the relative actions ${\Delta}{\cal F}_{ \overline{0} \gamma_1\gamma_2 \overline{0} ,  \overline{0} \gamma_1 \overline{0} } $ and ${\Delta}{\cal F}_{ \overline{0} \gamma_2\gamma_1 \overline{0} ,  \overline{0} \gamma_2 \overline{0} }$, which, following Eq.~\eqref{eq:homoclinic action difference2}, can be expressed as
\begin{equation}\label{eq:Auxiliary homoclinic relative actions 1 appendix}
\begin{split}
& {\Delta}{\cal F}_{ \overline{0} \gamma_1\gamma_2 \overline{0} ,  \overline{0} \gamma_1 \overline{0} } \\
& = \lim_{N \to \infty} \left[ F\left( h^{ ( \gamma_1\gamma_2 ) }_{ -( N+n^{+}_1 ) } , h^{ ( \gamma_1\gamma_2 ) }_{ n^{-}_2 +N  } \right) - F \left( h^{ ( \gamma_1 ) }_{ -N } , h^{ ( \gamma_1 ) }_{ N }  \right) \right]\\
&\quad - (n^{+}_1 + n^{-}_2) {\cal F}_0 
\end{split}
\end{equation}
and 
\begin{equation}\label{eq:Auxiliary homoclinic relative actions 2 appendix}
\begin{split}
& {\Delta}{\cal F}_{ \overline{0} \gamma_2\gamma_1 \overline{0} ,  \overline{0} \gamma_2 \overline{0} } \\
& = \lim_{N \to \infty} \left[ F\left( h^{ ( \gamma_2\gamma_1 ) }_{ -( N+n^{+}_2 ) } , h^{ ( \gamma_2\gamma_1 ) }_{ n^{-}_1 +N  } \right) - F \left( h^{ ( \gamma_2 ) }_{ -N } , h^{ ( \gamma_2 ) }_{ N }  \right) \right]\\
&\quad - (n^{+}_2 + n^{-}_1) {\cal F}_0 \ .
\end{split}
\end{equation}

Similar to Eq.~\eqref{eq:Difference between periodic and homoclinic relative actions entire string}, comparing ${\cal F}_{\gamma}$ to the sum of the relative auxiliary homoclinic orbit actions $( {\Delta}{\cal F}_{ \overline{0} \gamma_1\gamma_2 \overline{0} ,  \overline{0} \gamma_1 \overline{0} }  +  {\Delta}{\cal F}_{ \overline{0} \gamma_2\gamma_1 \overline{0} ,  \overline{0} \gamma_2 \overline{0} })$ gives
\begin{equation}\label{eq:Difference between periodic and homoclinic relative actions partitioned string}
\begin{split}
&{\cal F}_{\gamma} - {\Delta}{\cal F}_{ \overline{0} \gamma_1\gamma_2 \overline{0} ,  \overline{0} \gamma_1 \overline{0} } - {\Delta}{\cal F}_{ \overline{0} \gamma_2\gamma_1 \overline{0} ,  \overline{0} \gamma_2 \overline{0} }\\
&= F \left( y_0, y_{ n^{+}_1 + n^{-}_2 } \right) + F \left ( y_{ n^{+}_1 + n^{-}_2 } , y_{ n_{\gamma}} \right)\\
&\quad - \lim_{N\to\infty} \Bigg[ F\left( h^{ ( \gamma_1\gamma_2 ) }_{ -( N+n^{+}_1 ) } , h^{ ( \gamma_1\gamma_2 ) }_{ n^{-}_2 +N  } \right) - F \left( h^{ ( \gamma_1 ) }_{ -N } , h^{ ( \gamma_1 ) }_{ N }  \right)\\
&\qquad \qquad + F\left( h^{ ( \gamma_2\gamma_1 ) }_{ -( N+n^{+}_2 ) } , h^{ ( \gamma_2\gamma_1 ) }_{ n^{-}_1 +N  } \right) - F \left( h^{ ( \gamma_2 ) }_{ -N } , h^{ ( \gamma_2 ) }_{ N }  \right)  \Bigg]\\
&\quad + n_{\gamma}{\cal F}_0 \ .
\end{split}
\end{equation}

Analogous to the spirit of Sec.~\ref{Action formula with entire periodic orbit strings}, partition the homoclinic orbit generating functions $F(h^{ ( \gamma_i ) }_{ -N } , h^{ ( \gamma_i ) }_{ N } )$ ($i=1,2$) into two parts, $F(h^{ ( \gamma_i ) }_{ -N } , h^{ ( \gamma_i ) }_{ 0 })$ and $F(h^{ ( \gamma_i ) }_{ 0 } , h^{ ( \gamma_i ) }_{ N })$, which correspond to the initial and final parts of the $\lbrace h^{(\gamma_i)}_0 \rbrace$ orbit segment, respectively. Also, partition the generating functions $F\left( h^{ ( \gamma_i\gamma_j ) }_{ -( N+n^{+}_i ) } , h^{ ( \gamma_i\gamma_j ) }_{ n^{-}_j +N  } \right)$ ($i,j=1,2$) into three parts, $F\left( h^{ ( \gamma_i\gamma_j ) }_{ -( N+n^{+}_i ) }\ ,\ h^{ ( \gamma_i\gamma_j ) }_{ -n^{+}_i  } \right)$, $F\left( h^{ ( \gamma_i\gamma_j ) }_{ -n^{+}_i }\ ,\ h^{ ( \gamma_i\gamma_j ) }_{ n^{-}_j } \right)$ and $F\left( h^{ ( \gamma_i\gamma_j ) }_{ n^{-}_j }\ ,\ h^{ ( \gamma_i\gamma_j ) }_{ n^{-}_j +N  } \right)$, that correspond to the initial, middle and final parts of $\lbrace h^{(\gamma_i\gamma_j)}_0\rbrace$ orbit segment, respectively. The expression for ${\cal F}_{\gamma}$ is obtained from substituting these generating function into Eq.~\eqref{eq:Difference between periodic and homoclinic relative actions partitioned string}, and regrouping them into action differences according similarity in the symbolic code sequences along their orbit segments:
\begin{equation}\label{eq:Difference between periodic and homoclinic relative actions partitioned string regrouping}
\begin{split}
&{\cal F}_{\gamma} - {\Delta}{\cal F}_{ \overline{0} \gamma_1\gamma_2 \overline{0} ,  \overline{0} \gamma_1 \overline{0} } - {\Delta}{\cal F}_{ \overline{0} \gamma_2\gamma_1 \overline{0} ,  \overline{0} \gamma_2 \overline{0} }\\
&=\lim_{N\to\infty} \left[ F\left( h^{ ( \gamma_1 ) }_{ -N }\ ,\ h^{ ( \gamma_1 ) }_{ 0 } \right) - F\left( h^{ ( \gamma_1\gamma_2 ) }_{ -( N+n^{+}_1 ) }\ ,\ h^{ ( \gamma_1\gamma_2 ) }_{ -n^{+}_1  } \right) \right]\\
&+ \left[ F\left( y_0\ ,\ y_{ n^{+}_1 + n^{-}_2 } \right) - F\left( h^{ ( \gamma_1\gamma_2 ) }_{ -n^{+}_1 }\ ,\ h^{ ( \gamma_1\gamma_2 ) }_{ n^{-}_2 } \right) \right]\\
&+\lim_{N\to\infty} \left[ F\left( h^{ ( \gamma_2 ) }_{ 0 }\ ,\ h^{ ( \gamma_2 ) }_{ N } \right) - F\left( h^{ ( \gamma_1\gamma_2 ) }_{ n^{-}_2 }\ ,\ h^{ ( \gamma_1\gamma_2 ) }_{ n^{-}_2 +N  } \right) \right]\\
&+\lim_{N\to\infty} \left[ F\left( h^{ ( \gamma_2 ) }_{ -N }\ ,\ h^{ ( \gamma_2 ) }_{ 0 } \right)  - F\left( h^{ ( \gamma_2\gamma_1 ) }_{ -( N+n^{+}_2 ) }\ ,\ h^{ ( \gamma_2\gamma_1 ) }_{ -n^{+}_2  } \right) \right]\\
&+ \left[ F\left( y_{ n^{+}_1 + n^{-}_2  }\ ,\ y_{ n_{\gamma} } \right) - F\left( h^{ ( \gamma_2\gamma_1 ) }_{ -n^{+}_2 }\ ,\ h^{ ( \gamma_2\gamma_1 ) }_{ n^{-}_1 } \right) \right]\\
&+\lim_{N\to\infty} \left[ F\left( h^{ ( \gamma_1 ) }_{ 0 }\ ,\ h^{ ( \gamma_1 ) }_{ N } \right) - F\left( h^{ ( \gamma_2\gamma_1 ) }_{ n^{-}_1 }\ ,\ h^{ ( \gamma_2\gamma_1 ) }_{ n^{-}_1 + N  } \right) \right]\\
&+ n_{\gamma}{\cal F}_0 \ .
\end{split}
\end{equation}
Notice that the first six terms in the above expression are differences in the generating functions between orbits segments regrouped according to similarities in the symbolic code sequences. Following the same procedures as Sec.~\ref{Action formula with entire periodic orbit strings} by repeated use of Eq.~\eqref{eq:Meiss92}, converts each term into a phase space integral along certain paths:

\begin{itemize}

\item[i)] the first term 
\begin{equation}\label{eq:Derivation step 1}
\begin{split}
& \lim_{N\to\infty} \left[ F\left( h^{ ( \gamma_1 ) }_{ -N }\ ,\ h^{ ( \gamma_1 ) }_{ 0 } \right) - F\left( h^{ ( \gamma_1\gamma_2 ) }_{ -( N+n^{+}_1 ) }\ ,\ h^{ ( \gamma_1\gamma_2 ) }_{ -n^{+}_1  } \right) \right]\\
&= \int\limits_{U[ h^{ ( \gamma_1\gamma_2 ) }_{ -n^{+}_1  }\ ,\ h^{ ( \gamma_1 ) }_{ 0 } ]} p\mathrm{d}q 
\end{split}
\end{equation}

\item[ii)] the second term, 
\begin{equation}\label{eq:Derivation step 2}
\begin{split}
&F\left( y_0\ ,\ y_{ n^{+}_1 + n^{-}_2 } \right) - F\left( h^{ ( \gamma_1\gamma_2 ) }_{ -n^{+}_1 }\ ,\ h^{ ( \gamma_1\gamma_2 ) }_{ n^{-}_2 } \right) \\
&= \int\limits_{C^{\prime}[ h^{ ( \gamma_1\gamma_2 ) }_{ n^{-}_2 }\ ,\ y_{ n^{+}_1 + n^{-}_2 } ]} p\mathrm{d}q + \int\limits_{C[ y_{ 0 }\ ,\ h^{ ( \gamma_1\gamma_2 ) }_{ -n^{+}_1 } ]} p\mathrm{d}q 
\end{split}
\end{equation}
where, similar to Sec.~\ref{Action formula with entire periodic orbit strings}, we take the liberty to choose $C$ to be the straight-line segment connecting the end points, which is mapped to a near straight-line segment $C^{\prime}$ under $(n^{+}_1 + n^{-}_2)$ iterations. 

\item[iii)] the third term 
\begin{equation}\label{eq:Derivation step 3}
\begin{split}
& \lim_{N\to\infty} \left[ F\left( h^{ ( \gamma_2 ) }_{ 0 }\ ,\ h^{ ( \gamma_2 ) }_{ N } \right) - F\left( h^{ ( \gamma_1\gamma_2 ) }_{ n^{-}_2 }\ ,\ h^{ ( \gamma_1\gamma_2 ) }_{ n^{-}_2 +N  } \right) \right] \\
&=\int\limits_{S[ h^{ ( \gamma_2 ) }_{ 0 }\ ,\ h^{ ( \gamma_1\gamma_2 ) }_{ n^{-}_2 } ]} p\mathrm{d}q 
\end{split}
\end{equation}

\item[iv)] the fourth term
\begin{equation}\label{eq:Derivation step 4}
\begin{split}
& \lim_{N\to\infty} \left[ F\left( h^{ ( \gamma_2 ) }_{ -N }\ ,\ h^{ ( \gamma_2 ) }_{ 0 } \right)  - F\left( h^{ ( \gamma_2\gamma_1 ) }_{ -( N+n^{+}_2 ) }\ ,\ h^{ ( \gamma_2\gamma_1 ) }_{ -n^{+}_2  } \right) \right] \\
&= \int\limits_{U[ h^{ ( \gamma_2\gamma_1 ) }_{ -n^{+}_2  } \ ,\ h^{ ( \gamma_2 ) }_{ 0 } ]} p\mathrm{d}q 
\end{split}
\end{equation}

\item[v)] the fifth term
\begin{equation}\label{eq:Derivation step 5}
\begin{split}
&\left[ F\left( y_{ n^{+}_1 + n^{-}_2  }\ ,\ y_{ n_{\gamma} } \right) - F\left( h^{ ( \gamma_2\gamma_1 ) }_{ -n^{+}_2 }\ ,\ h^{ ( \gamma_2\gamma_1 ) }_{ n^{-}_1 } \right) \right]\\
&= \int\limits_{C^{\prime}[ h^{ ( \gamma_2\gamma_1 ) }_{ n^{-}_1 } \ ,\ y_{ n_{\gamma} } ]} p\mathrm{d}q + \int\limits_{C[ y_{ n^{+}_1 + n^{-}_2 } \ ,\ h^{ ( \gamma_2\gamma_1 ) }_{ -n^{+}_2 } ]} p\mathrm{d}q 
\end{split}
\end{equation}
where $C$ is chosen to be the straight-line segment connecting the end points, which is mapped to a near straight-line segment $C^{\prime}$ under $(n^{-}_1 + n^{+}_2)$ iterations. 

\item[vi)] the sixth term
\begin{equation}\label{eq:Derivation step 6}
\begin{split}
& \lim_{N\to\infty} \left[ F\left( h^{ ( \gamma_1 ) }_{ 0 }\ ,\ h^{ ( \gamma_1 ) }_{ N } \right) - F\left( h^{ ( \gamma_2\gamma_1 ) }_{ n^{-}_1 }\ ,\ h^{ ( \gamma_2\gamma_1 ) }_{ n^{-}_1 + N  } \right) \right]\\
&= \int\limits_{S[ h^{ ( \gamma_1 ) }_{ 0 }\ ,\ h^{ ( \gamma_2\gamma_1 ) }_{ n^{-}_1 } ]} p\mathrm{d}q \ .
\end{split}
\end{equation}
\end{itemize}

Substituting Eqs.~\eqref{eq:Derivation step 1} - \eqref{eq:Derivation step 6} into Eq.~\eqref{eq:Difference between periodic and homoclinic relative actions partitioned string regrouping} and rearranging the terms eventually leads to
\begin{equation}\label{eq:Homoclinic decomposition action formula reshuffle exact letter form appendix}
\begin{split}
{\cal F}_{\gamma} &= n_{\gamma}{\cal F}_{0} + {\Delta}{\cal F}_{ \overline{0} \gamma_1\gamma_2 \overline{0} ,  \overline{0} \gamma_1 \overline{0} }  +  {\Delta}{\cal F}_{ \overline{0} \gamma_2\gamma_1 \overline{0} ,  \overline{0} \gamma_2 \overline{0} }\\
& + {\cal A}^\circ_{ C U S C^{\prime} [ y_0\ ,\ h^{ (\gamma_1\gamma_2) }_{ -n^{+}_1 }\ ,\ h^{ (\gamma_1) }_0\ ,\ h^{ (\gamma_2\gamma_1) }_{ n^{-}_1 } ] } \\
& + {\cal A}^\circ_{ C U S C^{\prime} [ y_{ n^{+}_1 + n^{-}_2 }\ ,\ h^{ (\gamma_2\gamma_1) }_{ -n^{+}_2 }\ ,\ h^{ (\gamma_2) }_0\ ,\ h^{ (\gamma_1\gamma_2) }_{ n^{-}_2 } ] } 
 \end{split}
\end{equation}
which is Eq.~\eqref{eq:Homoclinic decomposition action formula reshuffle exact} in Sec.~\ref{Action formula with reshuffled substrings}.

\bibliography{classicalchaos,quantumchaos}

\end{document}